\documentclass[10pt]{article}
\usepackage{latexsym}
\usepackage[dvips]{graphicx}
\usepackage{overcite}
\usepackage{amssymb}
\usepackage{amsbsy}
\usepackage{amsmath}
\usepackage{mathrsfs}
\usepackage{xr}
\usepackage{booktabs} 
\usepackage{multirow} 
\usepackage{xcolor}	  
\usepackage{mathtools}
\usepackage{graphicx}
\usepackage{tabularx}
\usepackage[labelformat=simple, font=small,labelfont=bf]{caption}
\usepackage{placeins}
\usepackage{authblk}
\graphicspath{ {./plot_reorga/} }

\DeclareRobustCommand*{\Figure}[3]{
   \begin{figure}[!htb]
   \begin{center}
   \noindent
   \includegraphics[width=#2]{#1}  
   \end{center}
   \caption{#3}
   \addtocontents{lof}{\vspace{\baselineskip}}
   \label{fig:#1}
   \end{figure}
}

\newcommand{\be}{\begin{equation}}
\newcommand{\ee}{\end{equation}}

\begin{document}
\title{Investigation of Polymer Association Behaviors in Solvents Using a Coarse-Grained Model}

\author[1]{Xiangyu Zhang\footnote{E-mail: xzhan357@jh.edu}}
\author[2]{Dong Meng}
\affil[1]{Department of Chemical and Biomolecular Engineering, John Hopkins University, Baltimore, Maryland 21218, United State}
\affil[2]{Biomaterials Division, Department of Molecular Pathobiology, New York University, New York, NY 10010, United State}

\maketitle

\begin{abstract}
The associative interaction, such as hydrogen bonding, can bring about versatile functionalities to polymer systems, which has been investigated by tremendous researches, but the fundamental understanding on association process is still lacking. In this study, a reaction-controlled association model is proposed to delve into the polymer association activities in solvents, which is proved to obey the principle of thermodynamics. Additionally, associative polymer chain configurational bias method is developed to improve sampling efficiency, demonstrating a significantly faster relaxation process. First, we set non-bonded interactions to be zero, and only keep the chain connectivity and association. It is found that the association process intrinsically follows Bernoulli process by comparing the simulation results and analytic results. Next, we include non-bonded interactions into the simulation to examine its effects. It emerged that the excluded volume effect and solvents immiscibility effects can result in inhomogeneous associating probability distribution along the chain contour, in contrast to the homogeneity observed in ideal systems, thereby shifting from the binomial distribution to Poisson binomial distribution. At last, the study is extended to cooperative association systems. The incorporation of cooperative association can lead to the coexistence of coil and globule state at the transition point, verified by the potential of mean force calculation. Finally, a mathematical model is proposed, illustrating the changes in statistical weight induced by sequence enthalpy bias, which is the consequence of cooperative behaviors.

\end{abstract}

\section{Introduction}
Reversible associations, such as hydrogen bonds, $\pi-\pi$ conjugation, metal-ligand coordination, ionic interactions, etc., provide versatile applications for polymer materials. \cite{song2020high, neo2016conjugated, he2015nanomedicine, potaufeux2020comprehensive} Notably, the introduction of association often brings about intriguing and complicated behaviors, in both polymer solution and polymer melt systems, resulted by the diverse and erratic association patterns. In polymer solution, hydrogen bond forming or breaking can lead to lower critical solution temperature transition, contrasted with upper critical solution temperature transition, which is $\chi$-interaction driven. \cite{zhang2015polymers, zhang2017thermoresponsive} Besides, polymer/solvents/cosolvents ternary mixture may exhibit some counter-intuitive phenomena, such as cosolvency, in which the association plays a significant role.\cite{Zhang2025Unraveling} In polymer melts, the reversibility of chemical bonds can give rise to complicated self-assembly pathways, accompanied with kinetically trapped meta-stable state. \cite{yan2016kinetic, evans2018self} Therefore, the development of a facile method to describe association correctly is needed. \par

Several strategies have been proposed to incorporate reversible associations into simulations. R. S. Hoy and G. H. Fredrickson added Monte Carlo (MC) bond forming/breaking movement to molecular dynamics (MD) simulation,  providing insights for dynamics and mechanical properties. \cite{hoy2009thermoreversible} The bond forming/breaking probability depends on the deviation of bond length from equilibrium bond length and sticky binding energy, which is an input parameter to adjust thermodynamics of the bond. \cite{hoy2009thermoreversible} But the constraints implemented in the algorithm defies the natural association process and it does not account for number of associating candidate effects. The model used by S. Wang \textit{et al.} draw the bonding configuration from Boltzmann probability distribution, which is calculated by using pre-defined energy change, but it does not account for the candidate's distance when selecting a bonding partner. \cite{wang2008reversible} S. Liu and T. C. O'Connor developed a model to introduce Tersoff bond potential into MD simulations. The advantage of it is that the association can come naturally from exploring free energy landscape without any restraints or imposed MC movements. However, the Tersoff bond potential needs adjustments to do tremendous parameter fitting, limiting the application in diverse systems. Besides, the kinetic trap caused by strong association is difficult to overcome in MD simulation. There are also reversible bond models to deal with bond rearrangement, like bond swapping method \cite{sciortino2017three,rovigatti2018self}, but it cannot describe the number of bonding partner fluctuations. 
K. Ch. Daoulas \textit{et al.} sampled chain connectivity matrices by performing bond forming/breaking moves,\cite{daoulas2009phase} drawing on the similar concepts from previous studies,\cite{wittmer1998dynamical,milchev2000dynamical,chen2004ring,chen2006monte} 
 though this approach lacks the flexibility to control the equilibrium point of the association process. Therefore, a straightforward yet rigorous model grounded in fundamental thermodynamic principles is essential for studying association activities at the coarse-grained level. Such a model should capture two most important characteristics of the association process: reversibility and topological correlation. \par  


Besides for the method to deal with normal association, the other point worth to mention is that how to implement association "variants" based on a normal association model. One of association "variants" examples is cooperative association, as it has been discussed in several works. Tanaka \textit{et al.} proposed a cooperative association model in the theoretical study, and it concludes that the correlation between bounded water molecules has been confirmed as the origin for the flat spinodal curve, and the sharp coil-globule transition results from the cooperativity. \cite{okada2005cooperative} Moreover, the proposed pearl-necklace conformation, in which sequential consecutive hydrogen bonds are presented and interrupted by dehydrated sequences, can be verified by neutron scattering method and nanofishing experiments. \cite{koizumi2004concentration, liang2018nanofishing, koizumi2019necklace} 
In addition to findings suggesting cooperative hydration — where an associated neighbor encourages association — experimental and detailed simulation studies also report cooperative dehydration, indicating that a non-associated neighbor promotes de-association. \cite{futscher2017role, kang2016collapse, liu2022coil}. Moreover, the cooperative behavior may exist in various types of systems, as it has been proposed that it may play an important role in protein folding-unfolding process, signifying the potential application in biomaterial rational design.\cite{huggins2016studying, miyawaki2016cooperative}. However, the direct method to incorporate association "variants" in simulation is still lacking.\par


In this study, we used a reaction-controlled association method to investigate association effect on single chain polymer conformation transition. The manuscript is organized as following. First, details about how to incorporate association is described in model and method section. Besides, we also develop associative polymer chain configurational bias (APCCB) to help relax the polymer chain, the detail of which is included in method section. In the result section, we first study the pure association process without any non-bonded interaction effects, investigating the nature of association process. The polymer chain relaxation result shows that APCCB method can greatly help improve sampling efficiency. Second, chain conformation transition induced by normal association is investigated, highlighting the non-bonded interaction effects on association. Third, we extend the normal association system to cooperative association system, showing the significant difference from normal association system resulted by cooperativity. Finally, a mathematical model was developed, and its results show consistency with the simulation outcomes.

\section{Model and Method}
\subsection{Model}
The system is consisted of single linear polymer chain with chain length equal to $N_{P}$ immersed in $n_{S}^{T}$ solvents with unit length, where polymer segments can form associative bonds with solvents. The simulation is running in NVT ensemble.  \par
The Hamiltonian of the system is given by:
\begin{equation}
\mathcal{H}=\mathcal{H}^{b}+\mathcal{H}^{nb}+\mathcal{H}^{a}
\end{equation}
where $\mathcal{H}^{b}$, $\mathcal{H}^{nb}$ and $\mathcal{H}^{a}$ are contributions from covalent bonds, non-bonded and associative bonds energy, respectively. Covalent bond energy is defined as, 
\begin{equation}
\mathcal{H}^{b}=\sum_{j=1}^{N_{P}-1}u^{b}(\left | {\bf r}_{P,j+1} - {\bf r}_{P,j} \right |)
\end{equation}
, where ${\bf r}_{P,j}$ denotes the spatial position of $j$ polymer segment, and $u^{b}$ is the bonding potential, retaining connectivity. In the study, discrete Gaussian bond potential is used,
\begin{equation} \label{eq:DGB_P}
u^{b}(\left | {\bf r}_{j+1} - {\bf r}_{j} \right |)=\frac{3 k_{B}T}{2a^{2}}\left | {\bf r}_{P,j+1} - {\bf r}_{P,j} \right |^{2}
\end{equation}
, with $a$ being the effective bond length, $k_{B}$ being the Boltzmann constant, and $T$ being the temperature. The total non-bonded energy is given by,
\begin{equation}
\begin{split}
\mathcal{H}^{nb} & = \sum_{i=1}^{N_P}\sum_{j>i}^{N_P}\int d{\bf r}\int d{\bf r}' \delta({\bf r}-{\bf r}_{P,i}) u_{PP}^{nb}({\bf r},{\bf r}')\delta({\bf r}'-{\bf r}_{P,j}) \\
& + \sum_{i=1}^{n_{S}^{T}}\sum_{j>i}^{n_{S}^{T}}\int d{\bf r}\int d{\bf r}' \delta({\bf r}-{\bf r}_{S,i}) u_{SS}^{nb}({\bf r},{\bf r}')\delta({\bf r}'-{\bf r}_{S,j}) \\
& +\sum_{i=1}^{N_P}\sum_{j=1}^{n_{S}^{T}}\int d{\bf r}\int d{\bf r}' \delta({\bf r}-{\bf r}_{P,i}) u_{PS}^{nb}({\bf r},{\bf r}')\delta({\bf r}'-{\bf r}_{S,j})
\end{split}
\end{equation}
, where ${\bf r}_{\alpha,k}$ represents the spatial position of the $\alpha$ type $k$ segment, and
\begin{equation}
u_{\alpha \alpha'}^{nb}({\bf r},{\bf r}') \equiv \epsilon_{\alpha\alpha'} \left( 15/2\pi \right) \left(1 - r/\sigma \right)^2
\end{equation}
for $r<\sigma$, where $\sigma$ is the unit length, or $u_{\alpha \alpha'}^{nb}({\bf r},{\bf r}') = 0$ otherwise. $\epsilon_{\alpha\alpha'}$ controls the interaction strength and has the unit of $k_{B}T$, which is defined as, 
$\epsilon_{\alpha\alpha\prime}\equiv\left\{\def\arraystretch{1.2}\begin{tabular}{@{}l@{\quad}l@{}}
  $\epsilon_{\kappa}$ & if $\alpha=\alpha\prime$ \\
  $\epsilon_{\kappa}+\epsilon_{\chi_{PS}}$ & if $\alpha \neq \alpha\prime$
\end{tabular}\right.$ . $\epsilon_{\kappa}$ is the excluded volume and $\epsilon_{\chi_{PS}}$ describes the solvent immiscibility, representing the hydrophobic interaction strength. $\epsilon_{\kappa}$ and $\epsilon_{\chi_{PS}}$ are both constants in the study, equal to $0.5$ and $3$, respectively. \par
The total association energy is the sum of the energy of all associative bonds, which can be expressed as,
\begin{equation}
\label{eq: asso_total}
\mathcal{H}^{a}=\sum_{j=1}^{N_{asso}}u^{a}(\left | {\bf r}_{P,j} - {\bf r}_{S,j} \right |)) .
\end{equation}
The next question becomes how to correctly capture the association behaviors in the polymer/solvents mixture.

\subsection{Coarse-Grained Association Model} 
To introduce association process into the simulation, we consider a association reaction $P+S \rightleftarrows PS$ between a monomer (P) and a solvent molecule (S) with an association constant of $K_{\alpha}\equiv \frac{z_{PS}}{z_{P}z_{S}}$, where $z_{P}$, $z_{S}$ and $z_{PS}$ are the single molecular activity of a monomer, a solvent and an association complex molecule, respectively. $z_{\alpha}$ is related to the component chemical potential in the mixture solution through the definition $\mathcal{Z}_{\alpha}\equiv e^{\beta(\mu_{\alpha}-g_{\alpha}^{0})}$, where $g_{\alpha}^{0}$ is the single-molecular Gibbs free energy of an $\alpha$-type molecule ($\alpha=P,S,PS$) in its reference state. By choosing the pure and ideal state as the reference state, there is
\begin{equation}
\beta g_{\alpha}^{0} = -\ln{\mathcal{Z}_{\alpha}^{id}}
\end{equation}
where the ideal-state molecular partition function is given by $\mathcal{Z}_{\alpha}^{id}=\frac{Q_{\alpha}^{id}}{\lambda_{T}^{3 N_{\alpha}}}$, with $\lambda_{T}$ being the de Broglie wave length, $M_{\alpha}$ being the molecular weight for each $\alpha$-type molecule, and $Q_{\alpha}^{id}$ being the conformational partition function of an $\alpha$-type molecule. Upon substitution, one obtains
\begin{equation}
K_{\alpha} \equiv \frac{z_{PS}}{z_{P}z_{S}} = \frac{e^{\beta(\mu_{PS}-g_{PS}^{0})}}{e^{\beta(\mu_{P}-g_{P}^{0})} e^{\beta(\mu_{S}-g_{S}^{0})}} = e^{\beta(\mu_{PS}-\mu_{P}-\mu_{S})} \frac{\lambda_{T}^{3M_{P}} \lambda_{T}^{3M_{S}}}{\lambda_{T}^{3M_{PS}}} \frac{Q_{PS}^{id}}{Q_{P}^{id} Q_{S}^{id}} = \frac{Q_{PS}^{id}}{Q_{P}^{id} Q_{S}^{id}}
\end{equation}
The equilibrium condition $\mu_{PS}=\mu_{P}+\mu_{S}$, meaning that the work of inserting a PS complexation pair into the mixture equals to the work of inserting one P segment and one S segment separately, and $M_{PS}=M_{P}+M_{S}$ had been used. $\frac{Q_{PS}^{id}}{Q_{P}^{id} Q_{S}^{id}} \equiv Q_{a}^{id}$ accounts for the conformational changes of P-S complex from the original molecules due to association complexation. While the value of $Q_{a}^{id}$ is given by $Q_{a}^{id}=K_{a}$, the specific integral form of $Q_{a}^{id}$ depends on the chemistry details of P-S association which is beyond the resolution at the coarse-grained level. Instead, the integral form of $Q_{b}^{id}$ offers a degree of freedom in the construction of a coarse-grained association model. We model $Q_{b}^{id}$ through introducing an effective "association bond" between the participating polymer and solvent segment, with the bond energy $u_{a}$ being a function of the separation between the center of mass of molecule $P({\bf r}_{P})$ and $S({\bf r}_{S})$, 
\begin{equation} \label{eq:K_a}
K_{a}=\int_{0}^{r_{cut}} e^{-\beta u^{a}({\bf r}_{P} - {\bf r}_{S})} d({\bf r}_{P} - {\bf r}_{S})
\end{equation}
, where $r_{cut}$ is the cut-off distance. While $u^{a}({\bf r}_{P} - {\bf r}_{S})$ implies that both separation and orientation of $P$ and $S$ segment would contribute to the bond energy, it does not necessarily represent a physical bond at this level of modeling. Therefore, one is free to choose the function form of $u^{a}({\bf r}_{P} - {\bf r}_{S})$. Out of convenience, we choose to use eq. \ref{eq:DGB_P} for bonding potential plus $h_{A}$, which is reaction equilibrium constant term,
\begin{equation}
u^{a}(\left | {\bf r}_{P} - {\bf r}_{S} \right |)) \equiv \frac{3 k_{B}T}{2\sigma^{2}}(\left | {\bf r}_{P} - {\bf r}_{S} \right |)^{2} + h_{A}
\end{equation}
By substituting $u^{a}$ into eq. \ref{eq:K_a}, we can obtain $h_{A}$ expression,
\begin{equation} \label{eq:h_{A}_0}
h_{A} = - \ln{K_{a}} + \ln{(\int_{0}^{r_{cut}} \exp{(\frac{3 k_{B}T}{2\sigma^{2}}(\left | {\bf r}_{P} - {\bf r}_{S} \right |)^{2})} d({\bf r}_{P} - {\bf r}_{S}))}
\end{equation}
The cut-off distance for association we choose is $r_{cut}=2\sigma$, so, the numerical value of volume integral is $3.008$. \par
The expression for $K_{a}$ can be derived analytically for an ideal system corresponding to an association-only scenario, 
\begin{equation}
K_{a}=\frac{[PS]}{[P][S]}=\frac{C_{e} N_{P}}{(1-C_{e})N_{P} [S]}
\end{equation}
where $[X]$ is the concentration of $X$ species, $C_{e}$ is the equilibrium conversion ratio of polymer segments, $N_{P}$ is the total number of polymer segments carrying associative sites. In single chain system, solvents are extremely excessive, so, $[S]$ is approximately $\rho_{0}$, where $\rho_{0}$ is overall solvents number density. Therefore, we have,
\begin{equation}
\rho_{0} K_{a} = \frac{C_{e}}{1-C_{e}}
\end{equation}
And by substituting it into eq. \ref{eq:h_{A}_0}, we can have the final $h_{A}$ and $C_{e}$ relation,
\begin{equation} \label{eq:h_above}
h_{A} = - \ln{\frac{C_{e}}{1-C_{e}}} + \ln{(3.008 \rho_{0})}
\end{equation}
, which can be considered as free energy difference between associated and unassociated states. If there is cooperative association, we can introduce additional term to control the cooperative strength. The association potential form $u^{a}(\left | {\bf r}_{P,j} - {\bf r}_{S,j} \right |))$ in eq. \ref{eq: asso_total} can be written as,
\begin{equation} \label{eq: asso_pair}
u^{a}(\left | {\bf r}_{P,j} - {\bf r}_{S,j} \right |)) = \frac{3 k_{B}T}{2\sigma^{2}}(\left | {\bf r}_{P} - {\bf r}_{S} \right |)^{2} + h_{A}(1+ C_{\delta h} \cdot s_{P,j})
\end{equation}
where $C_{\delta h}$ is the cooperative coefficient, of whose value being negative means that the associated segment promotes its neighbor's association tendency. If it is positive, the associated segment will decrease its neighbors' associating probability. $s_{P,j}$ describes the number of covalent-bonded neighbors associated with the solvent. For linear chain, it can only take three possible values, $0$, $1$ and $2$. If $C_{\delta h}$ is set to be zero, only $h_{A}$ term is left, corresponding to non-cooperative system or normal association system. Replica exchange is used in cooperative systems simulation, of which the detail is presented in appendix.

\subsection{Association Bond Forming/Breaking Protocol} 
The detailed balance condition of cooperative association can be written as following equation, 
\begin{equation}
T(B \rightarrow F)P(F)acc(B \rightarrow F) = T(F \rightarrow B)P(B)acc(F \rightarrow B)
\end{equation}
where $acc$ is the acceptance criterion, $P(F)$ or $P(B)$ is the probability to observe a formed hydrogen bond or a broken bond, $T(B \rightarrow F)$ or $T(F \rightarrow B)$ is the proposing transition probability from unassociated state to associated state or from associated state to unassociated state. First, a segment is picked randomly, and the bond forming/breaking trials depend on its associating states. If it is associated, it will attempt to break the bond, and vice versa.\\
(1) if $i^{\mathrm{th}}$ polymer segment was picked to form/break a bond with $j^{\mathrm{th}}$ solvent, the association state of its bonded neighbor along the chain (i.e. the $(i+1)^{\mathrm{th}}$ and $(i-1)^{\mathrm{th}}$ segment) can be described by $s_{P,i}$ that takes three possible values $0,1,2$, corresponding to the three situations in which none, one or both of its neighbors are associated. The probability to observe a bond forming is,
\begin{equation}
\frac{P(B)}{P(F)}=\exp(\Delta U(F))=\exp(- h_{A} - \delta h \cdot s_{P,i} - E^{a}(\left | {\bf r}_{P,j} - {\bf r}_{S,j} \right |))
\end{equation}
or the bond breaking is,
\begin{equation}
\frac{P(F)}{P(B)}=\exp(\Delta U(B))=\exp( h_{A} + \delta h \cdot s_{P,i} + E^{a}(\left | {\bf r}_{P,j} - {\bf r}_{S,j} \right |))
\end{equation}
where $\delta h$ is $C_{\delta h} \cdot h_{A}$. The derivation is using $\delta h$ for convenience.
The Rosenbluth weight factor of bond-forming with solvent $j$ is,
\begin{equation} \label{eq:w_j_equation}
w_{j}=\exp(\Delta U(F))
\end{equation}
If polymer segment $i$ is open, the bond forming trial with solvent $j$ can be proposed according to the probability,
\begin{equation}
T(B \rightarrow F)=\frac{w_{j}}{W} .
\end{equation} 
And the Rosenbluth weight is,
\begin{equation} \label{eq:W}
W=\sum_{j=1}^{N_{can}}w_{j}
\end{equation}
 , where $N_{can}$ is the number of bonding candidates. If polymer segment $i$ is associated, the bond breaking trial with solvent $j$ can be written as, $T(F \rightarrow B)=1$, because there is only one candidate to be broken. Substituting the above equations into detailed balance, the Metropolis acceptance criterion \cite{metropolis1953equation} for bond forming becomes,
\begin{equation}
\label{eqn:asso_BF}
acc(B \rightarrow F)=\min(1, \frac{T(F \rightarrow B)P(B)}{T(B \rightarrow F)P(F)})=\min(1,W) 
\end{equation}
the acceptance criterion for bond breaking becomes,
\begin{equation}
\label{eqn:asso_FB}
acc(F \rightarrow B)=\min(1, \frac{T(B \rightarrow F)P(F)}{T(F \rightarrow B)P(B)})=\min(1,\frac{1}{W})
\end{equation}

(2) If $j^{\mathrm{th}}$ solvent was picked to form/break a bond with the polymer segment $i$, the difference from picking a polymer segment is that $s_{P,i}$ depends on associating states of polymer segments' candidates. But in case (1), $s_{P,i}$ is a constant among all solvent association candidates. By following the process in case (1), the final acceptance criterion for bond forming/breaking shares the same form as eq. \ref{eqn:asso_BF} and \ref{eqn:asso_FB}.

(3) The above association protocol can be justified analytically in ideal system, in which we only consider covalent-bonded energy ($\mathcal{H}^{b}$) and association energy ($\mathcal{H}^{a}$). Assuming that the equilibrium conversion ratio of the system is $C_{e}$. It is known that solvents distribute homogeneously in ideal systems, so, we can have the number of associating solvents candidates at distance $r_{b}$, 
\begin{equation}
N_{can}=4\pi\rho_{0}r_{b}^{2}\delta r
\end{equation}
, where $r_{b}$ is the distance to the chosen polymer segment, and $\delta r$ is the shell thickness, which is approaching $0$. Accordingly, eq. \ref{eq:w_j_equation} can be expressed as a function of $r_{b}$,
\begin{equation}
w(r_{b})=4\pi\rho_{0}r_{b}^{2}\delta r \exp{(-\frac{3}{2}r_{b}^{2}-h_{A})} .
\end{equation}
Therefore, Rosenbluth weight, that is eq. \ref{eq:W}, can be written as 
\begin{equation}
W=4\pi\rho_{0}\int_{0}^{r_{cut}} r_{b}^{2} \exp{(-\frac{3}{2}r_{b}^{2}-h_{A})} d r_{b}.
\end{equation}
The integral part is a constant, which can be defined as 
\begin{equation}
\mathcal{Z}_{ab}\equiv 4\pi \int_{0}^{r_{cut}} r_{b}^{2} \exp{(-\frac{3}{2}r_{b}^{2})} d r_{b}
\end{equation}
 . And the numerical value of $\mathcal{Z}_{ab}$ is $3.008$ when $r_{cut}=2\sigma$. So, $W$ can be written as
\begin{equation} \label{eq:W_analytical}
W=\rho_{0} \exp{(-h_{A})} \mathcal{Z}_{ab} .
\end{equation}
The ideal equilibrium condition is $N_{f}P_{f}-N_{b}P_{b}=0$, where $N_{f/b}$ is denoted by the number of trials to form/break the bond, and $P_{f/b}$ is denoted by the probability to form or break a bond, meaning that the total number of associative bonds keeps unchanged. Next, we can write down following two equations,
\begin{equation}
\begin{aligned}
N_{f} &= N_{T}\cdot (1-C_{e})	\\
N_{b} &= N_{T}\cdot C_{e}
\end{aligned}
\end{equation}
, where $N_{T}=N_{f}+N_{b}$. By substituting it into equilibrium condition, we can obtain $\frac{P_{f}}{P_{f}+P_{b}}=C_{e}$. By using acceptance criterion, we know that $W=P_{f}$, and $1/W=P_{b}$, but one thing needed to mention is that we take $1$ for the value of $W$ or $1/W$ when they exceed $1$. At last, we can obtain the $h_{A}$ and $C_{e}$ relation by using eq. \ref{eq:W_analytical} derived above,
\begin{equation} \label{eq:h_below}
h_{A}=-\ln{\frac{C_{e}}{1-C_{e}}}+\ln{(\rho_{0} \mathcal{Z}_{ab})}
\end{equation}
, which is consistent with eq. \ref{eq:h_above}.

\subsection{Associative Polymer Chain Configuration Bias Method} 
To help relax the chain conformation with associated solvents, associative polymer chain configuration bias (APCCB) method, including both free-end and fixed-end bias, is developed and implemented. The derivation process uses some concepts from previous studies. \cite{harris1988lattice,frenkel1992novel,siepmann1992configurational,frenkel2023understanding} The derivation details for dry polymer chain configurational bias is shown in steps (1) $\sim$ (4). And APCCB is shown in step (5). The scheme is described as following. \\
(1) If there were $N_{P}$ segments on a polymer chain, two segments are randomly chosen, denoted as $i_{start}$ and $i_{end}$, which will be considered as the two fixed ends for regrowing. Number of segments to be regrown is $n=i_{end}-i_{start}-1$. And the number of regrowing bonds is ${n+1}$. For convenience, in following derivations, $(i_{start}+1)^{\mathrm{th}}$ segment is denoted as $1$, the first segment to be regrown, and $(i_{end}-1)^{\mathrm{th}}$ segment is denoted as $n$, that is the last segment to be regrown. \\
(2) First, Rosenbluth weight at "new state" is calculated. The proposed bond length in each dimension obeys 1D Gaussian distribution. And there are totally $k$ proposed trials for each segment. For $i^{\mathrm{th}}$ segment, the proposed probability is, 
\begin{equation}
q({\bf R}_{i,new})=\frac{\exp{[-u^{b}_{i}({\bf R}_{i,new})]}}{\int_{-\infty}^{+\infty} \exp[-u^{b}_{i}({\bf R}_{i})]d{\bf R}_{i}}
\end{equation} 
where ${\bf R}_{i,new}$ is the position vector of $i^{\mathrm{th}}$ segment in new states, and $u^{b}_{i}$ is the potential of bond between $i$ and $i-1$ segment. The denominator can be considered as a constant denoted by $C$.\\
The non-bonded potential of $i^{\mathrm{th}}$ segment at $l^{\mathrm{th}}$ trial can be denoted as $w_{i}({\bf R}_{i,l})$. After proposing all $k$ trails, the probability to choose "new state" from them is,
\begin{equation}
\mathcal{N}({\bf R}_{i,new})=\frac{\exp[-w_{i}({\bf R}_{i,new})-u^{b}_{i}({\bf R}_{i,new})]g({\bf R}_{i,new})}{C\sum_{l=1}^{k}\exp[-w_{i}({\bf R}_{i,l})]g({\bf R}_{i,l})}
\end{equation} 
where $g({\bf R}_{i,l})$ is the guiding probability of $i^{\mathrm{th}}$ segment at $l^{\mathrm{th}}$ trial. 
$g$ is the probability of random walk of an ideal chain given a fixed starting point to find the other fixed end in $N$ steps. And it simply has the following equation form as we are using discrete Gaussian bond model \cite{fredrickson2006equilibrium},
\begin{equation}
g({\bf R}_{i,l})=(\frac{3}{2 \pi N})^{\frac{3}{2}} \exp(- \frac{3|{\bf R}|^2}{2N})
\end{equation}
 , where $N$ is the number of steps taken to find the end point, and $|{\bf R}|$ is the distance from starting segment to end segment. \\
The $n^{\mathrm{th}}$ segment, that is the last segment to be regrown, has two bonds connected to it. So, the equation of choosing probability for the $n^{\mathrm{th}}$ segment should include the potential of the bond connecting $(i_{end}-1)^{\mathrm{th}}$ and $i_{end}^{\mathrm{th}}$ segment,
\begin{equation}
\mathcal{N}({\bf R}_{n,new})=\frac{\exp[-w_{n}({\bf R}_{n,new})-u^{b}_{i}({\bf R}_{n,new})]g({\bf R}_{n,new})\exp[-u^{b}_{n+1}({\bf R}_{n,new})]}{C\sum_{l=1}^{k}\exp[-w_{n}({\bf R}_{n,l})]g({\bf R}_{n,l})\exp[-u^{b}_{n+1}({\bf R}_{n,l})]}
\end{equation}
By organizing above equations, the Rosenbluth weight of new state can be written as,
\begin{equation}
\begin{split}
W(new)= & \frac{1}{k}  \cdot \{\prod_{i=1}^{n-1} \{ C\sum_{l=1}^{k}\exp[-w_{i}({\bf R}_{i,l})]g({\bf R}_{i,l}) \}  \\
& \{ C\sum_{l=1}^{k}\exp[-w_{n}({\bf R}_{n,l})] g({\bf R}_{n,l}) \exp[-u^{b}_{n+1}({\bf R}_{n,l})] \} \} 
\end{split}
\end{equation}
(3) The next step is to calculate Rosenbluth weight of "old state". With the old position of segments already known, $k-1$ trials need to be proposed. Similarly, Rosenbluth weight for "old" states can be written as,
\begin{equation}
\begin{split}
W(old)= & \frac{\prod_{i=1}^{n-1} \{ C\exp[-w_{i}({\bf R}_{i,old})]g({\bf R}_{i,old})+C\sum_{l=1}^{k-1}\exp[-w_{i}({\bf R}_{i,l})]g({\bf R}_{i,l}) \} }{k} \\
& \cdot \{ C\exp[-w_{n}({\bf R}_{n,old})]\exp[-u^{b}_{n+1}({\bf R}_{n,old})]g({\bf R}_{n,old})+ \\
& C\sum_{l=1}^{k-1}\exp[-w_{n}({\bf R}_{n,l})]\exp[-u^{b}_{n+1}({\bf R}_{n,l})]g({\bf R}_{n,l}) \}
\end{split}
\end{equation}
(4) Justification of Algorithm\\
From the above deviation, the probability to transit from trial positions to new configurations is,
\begin{equation}
T({old \rightarrow new})=\frac{k\prod_{i=1}^{n}\exp[-w_{i}({\bf R}_{i,new})-u^{b}_{i}({\bf R}_{i,new})]g({\bf R}_{i,new})\exp[-u^{b}_{n+1}({\bf R}_{n,new})]}{W(new)}
\end{equation} 
The probability to observe the new configuration is,
\begin{equation}
P(new)=\prod_{i=1}^{n}\exp[-w_{i}({\bf R}_{i,new})-u^{b}_{i}({\bf R}_{i,new})]\exp[-u^{b}_{n+1}({\bf R}_{n,new})]
\end{equation}
Similarly, $T_{new\rightarrow old}$ and $P(old)$ can be written as,
\begin{equation}
T({new \rightarrow old})=\frac{k\prod_{i=1}^{n}\exp[-w_{i}({\bf R}_{i,old})-u^{b}_{i}({\bf R}_{i,old})]g({\bf R}_{i,old})\exp[-u^{b}_{n+1}({\bf R}_{n,old})]}{W(old)}
\end{equation}
\begin{equation}
P(old)=\prod_{i=1}^{n}\exp[-w_{i}({\bf R}_{i,old})-u^{b}_{i}({\bf R}_{i,old})]\exp[-u^{b}_{n+1}({\bf R}_{n,old})]
\end{equation}
By imposing the detailed balance and substituting the derived equations into it, it can be found that only Rosenbluth weight and guiding probability are left.
\begin{equation}
\frac{acc({old \rightarrow new})}{acc({old \rightarrow new})}=\frac{P(new)T({new \rightarrow old})}{P(old)T({old \rightarrow new})}=\frac{W(new)G(old)}{W(old)G(new)}
\end{equation}
where $G(old(new))=\prod_{i=1}^{n}g({\bf R}_{i,old(new)})$.
Finally, Metropolis acceptance criterion is,
\begin{equation}
acc({old \rightarrow new})=\min(1,\frac{W(new)G(old)}{W(old)G(new)})
\end{equation}
(5) Configurational Bias Method with Associated Solvents \\
Assuming there are $n_{S}$ solvents associated with the polymer chain, the solvent regrowth process must be incorporated through the following modification. The proposing probability can be written as,
\begin{equation}
q({\bf R}_{i_{S},new})=\frac{\exp[-u^{a}_{i_{s}}({\bf R}_{i_{s},new})]}{\int_{-2}^{2}\exp[-u^{a}_{i_{s}}({\bf R}_{i_{s}})] d{\bf R}_{i_{s}}}
\end{equation}
where $i_{s}$ is the index of solvents to be regrown, $2$ is the truncated distance for associative bonds. The denominator is a constant, denoted as $C_{S}$.
The probability of finding the old and new state are,
\begin{equation}
\begin{split}
\frac{P(new)}{P(old)}= & \frac{\prod_{i=1}^{n}\exp[-w_{i}({\bf R}_{i,new})-u^{b}_{i}({\bf R}_{i,new})]\exp[-u^{b}_{n+1}({\bf R}_{n,new})]}{\prod_{i=1}^{n}\exp[-w_{i}({\bf R}_{i,old})-u^{b}_{i}({\bf R}_{i,old})]\exp[-u^{b}_{n+1}({\bf R}_{n,old})]}  \\
& \cdot \frac{\prod_{i_{s}=1}^{n_{S}}\exp[-u^{a}_{i_{s}}({\bf R}_{i_{s},new})-w_{i_{s}}({\bf R}_{i_{s},new})]}{\prod_{i_{s}=1}^{n_{S}}\exp[-u^{a}_{i_{s}}({\bf R}_{i_{s},old})-w_{i_{s}}({\bf R}_{i_{s},old})]}
\end{split}
\end{equation}
The transition probability for old and new state are,
\begin{equation}
\begin{split}
T({new \rightarrow old})= & \frac{1}{W(old)} \cdot \{ \prod_{i=1}^{n}\exp[-w_{i}({\bf R}_{i,old})-u^{b}_{i}({\bf R}_{i,old})]\exp[-u^{b}_{n+1}({\bf R}_{n,old})]   \\
& \cdot \prod_{i_{s}=1}^{n_{S}}\exp[-u^{a}_{i_{s}}({\bf R}_{i_{s},old})-w_{i_{s}}({\bf R}_{i_{s},old})] \}
\end{split}
\end{equation}
\begin{equation}
\begin{split}
T({old \rightarrow new})= & \frac{1}{W(new)} \cdot \{ \prod_{i=1}^{n}\exp[-w_{i}({\bf R}_{i,new})-u^{b}_{i}({\bf R}_{i,new})]\exp[-u^{b}_{n+1}({\bf R}_{n,new})] \\
& \cdot \prod_{i_{s}=1}^{n_{S}}\exp[-u^{a}_{i_{s}}({\bf R}_{i_{s},new})-w_{i_{s}}({\bf R}_{i_{s},new})] \}
\end{split}
\end{equation}
where $W(old)$ and $W(new)$ are,
\begin{equation}
\begin{split}
W(old)= & \frac{\prod_{i=1}^{n-1} \{ C\exp[-w_{i}({\bf R}_{i,old})]g({\bf R}_{i,old})+\sum_{l=1}^{k-1}\exp[-w_{i}({\bf R}_{i,l})]g({\bf R}_{i,l}) \} }{k} \\
& \cdot \{ C\exp[-w_{n}({\bf R}_{n,old})]\exp[-u^{b}_{n+1}({\bf R}_{n,old})]g({\bf R}_{n,old})+ \\
& \sum_{l=1}^{k-1}\exp[-w_{n}({\bf R}_{n,l})]\exp[-u^{b}_{n+1}({\bf R}_{n,l})]g({\bf R}_{n,l}) \} \\
& \cdot \prod_{i_{S}=1}^{n_{S}}C_{S}(\sum_{l=1}^{k-1}\exp[-w_{i_{S}}({\bf R}_{i_{S},l})]+\exp[-w_{i_{S}}({\bf R}_{i_{S},old})])
\end{split}
\end{equation}
\begin{equation}
\begin{split}
W(new)= & \frac{1}{k} \cdot \prod_{i=1}^{n-1} \{ C\sum_{l=1}^{k}\exp[-w_{i}({\bf R}_{i,l})]g({\bf R}_{i,l}) \} \prod_{i_{S}=1}^{n_{S}}C_{S}\sum_{l=1}^{k}\exp[-w_{i_{S}}({\bf R}_{i_{s},l})]  \\
& \cdot \{ C\sum_{l=1}^{k}\exp[-w_{n}({\bf R}_{n,l})] g({\bf R}_{n,l}) \exp[-u^{b}_{n+1}({\bf R}_{n,l})] \} 
\end{split}
\end{equation}
The final acceptance criterion can be written as,
\begin{equation}
acc({old \rightarrow new})=\min(1, \frac{W(new)G(old)}{W(old)G(new)})
\end{equation}
(6) Free End Configurational Bias\\
There are two key differences between the free-end configurational bias and the fixed-end bias. One is that the guiding probability is not needed, in other words, it can be considered as $1$. The other one is that the $u_{n+1}$ bonded energy term does not exist, as the last segment has only one bond connected to it.

\section{Results}
\subsection{The Pure Association Process} \label{sec:asso_nature}
We first examine the nature of pure association process for single chain immersed in solvents by only keeping the association process and chain connectivity. We define such systems as ideal systems. The excluded volume and immiscibility interactions are set to $0$ in the simulation ($\epsilon_{\kappa}$ and $\epsilon_{\chi}$ are set to be $0$), and the segment can still do spacial hopping movements. It has been proposed that the hydrogen bond sequences obey Bernoulli process \cite{dahanayake2021hydrogen}. Indeed, each segment is a two-state variable, to be associated or unassociated, making individual segment's associating activity a Bernoulli trail. To examine the associative activities in our model, we test the association process at $\rho_{0}=0.8 /\sigma^{3}$ with polymer chain length being equal to $200$. Figure~\ref{fig:figure_1.png} shows the distribution of the number of associated segments at three different conversion ratios: $0.8$, $0.46$, and $0.2$, and the conversion ratio is defined as the number of associated polymer segments divided by the total number of polymer segments carrying associative sites. x-axis is the number of associated segments, and y-axis is the corresponding probability. It can be found that all of distributions in ideal system almost perfectly overlap with the binomial distribution curve, indicating the Bernoulli process. To delve deeper, we know that the binomial probability can be calculated as,
\begin{equation}
P(m,n)={\binom{n}{m}}p^{m}(1-p)^{n-m}
\end{equation}
, where $n$ can be considered as total number of segments carrying associating sites, $m$ is the number of associated segments, and $p$ is the success probability of a single trial, same as average conversion rate when all of segments are indistinguishable. The above equation can be rearranged into exponential forms,
\begin{equation}
P(m,n) = {\binom{n}{m}}\exp{(m \ln{\frac{p}{1-p}})} \exp{(n \ln{(1-p)})} .
\end{equation}

Next, we can rewrite it into the following equivalent form to give a more intuitive expression, 
\begin{equation} \label{eq:asso_distribution}
\begin{aligned}
P(m,n) & = \frac{{\binom{n}{m}} \exp{(m \ln{(\frac{p}{1-p})})} \exp{(n \ln{(1-p)})}}{\exp{(n \ln{(1-p)})} \sum_{m=0}^{n} {\binom{n}{m}} \exp{(m \ln{(\frac{p}{1-p})})}} \\
& = \frac{{\binom{n}{m}} \exp{(m \ln{(\frac{p}{1-p})})} }{ \sum_{m=0}^{n} {\binom{n}{m}} \exp{(m \ln{(\frac{p}{1-p})})}} . 
\end{aligned}
\end{equation}
The denominator of first equation in eq. \ref{eq:asso_distribution} is equal to $1$, as it is the sum of all probabilities. It can be observed that the only variable affecting $P(m,n)$ is $\frac{p}{1-p}$ at a given $m$, so, we can propose the following relation between binomial distribution and association activity in our model if the association process exactly follows Bernoulli process,
\begin{equation} \label{eq:h_A_relation}
h_{A} \sim -\ln{(\frac{p}{1-p})} + C
\end{equation}
, where $h_{A}$ is the model parameter to adjust associating probability, as it is the only input parameter to change associating probability in our model, and $C$ is the shift constant. 

It can be seen that eq. \ref{eq:h_A_relation} has the same form as eq. \ref{eq:h_above} derived based on thermodynamic principles and eq. \ref{eq:h_below} derived based on simulation algorithm. 
Therefore, it can be concluded that the ideal system association activities universally is pure Bernoulli process, and it should obey binomial distribution. Moreover, the above relation can also be justified from the perspective of statistical mechanics. Supposing that the probability to observe associating state is $P_{\text{asso}}=\exp{(-a)}/\mathcal{Z}$ and there are no correlations, where $\mathcal{Z}$ is the partition function and $a$ is the free energy at associating state, and the probability for unassociating state is $P_{\text{un}}=\exp{(-b)}/\mathcal{Z}$, so, we can obtain $-\ln{(\frac{P_{\text{asso}}}{P_{\text{un}}})}=a-b$, where $P_{\text{asso}}$ is $C_{e}$, $P_{\text{un}}$ is $1-C_{e}$, and $h_{A}$ is effectively the $b-a$.  \par

Next, we test the $h_{A}$ and $C_{e}$ relation by using two systems with overall solvents number density being $1.2$ and $0.8$, respectively. In figure~\ref{fig:figure_2_new.png} (a), $h_{A}$ is plotted against average conversion value ($C_{e}$). Besides, analytic value calculated by eq. \ref{eq:h_below} is shown by dashed line, and they basically overlap with the simulation results. All systems should have the same slope at the same conversion point regardless of solvents density as is suggested by eq. \ref{eq:h_below}. $\rho_{0}=1.2 /\sigma^{3}$ requires higher association energy penalty to reach the same conversion value caused by the increase of the association candidate density. If we take the first order derivative of $h_{A}$ with respect to conversion rate to ignore $\rho_{0}$ and $\mathcal{Z}_{ab}$ shifting effect, we can get
\begin{equation} \label{eq:dh_dp}
\frac{d h_{A}}{d C_{e}} = \frac{1}{(C_{e}-1)C_{e}} . 
\end{equation}
We can conclude that the association process in our simulations is consistent from all of aspects, and the simulation results are consistent with analytic predictions, indicating the ability to model associative behaviors. In this section, all non-bonded interactions are screened out, as we focus on investigating the pure association process. The excluded volume and polymer-solvents immiscibility effect on transition curve will be discussed in section~\ref{section:normal_asso}. \par

\Figure{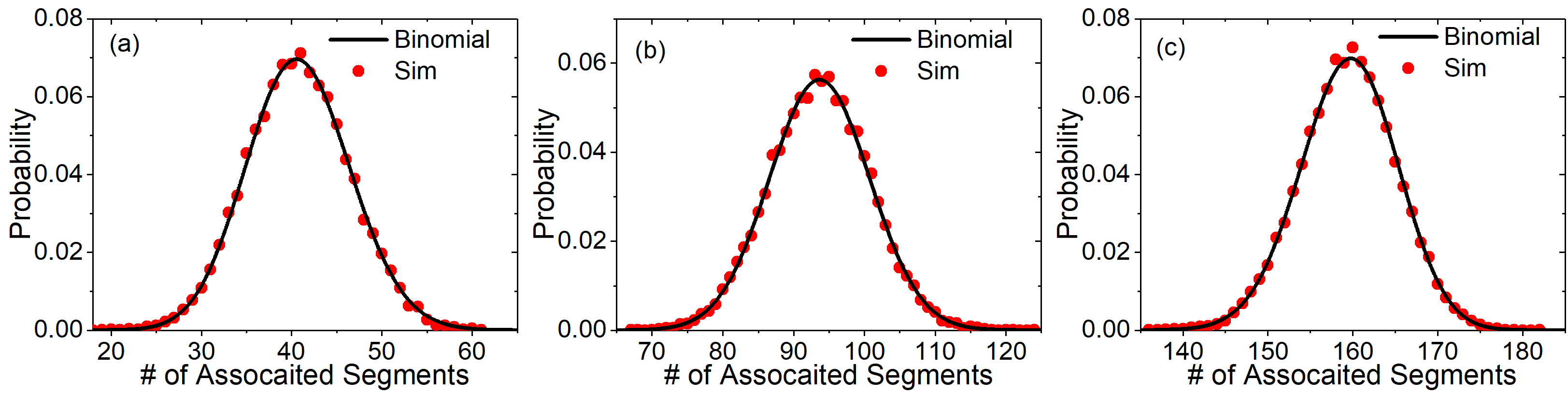}{0.98\linewidth}{The number of associated segments distribution for systems at the average conversion of $0.2$, $0.46$ and $0.8$.}
\Figure{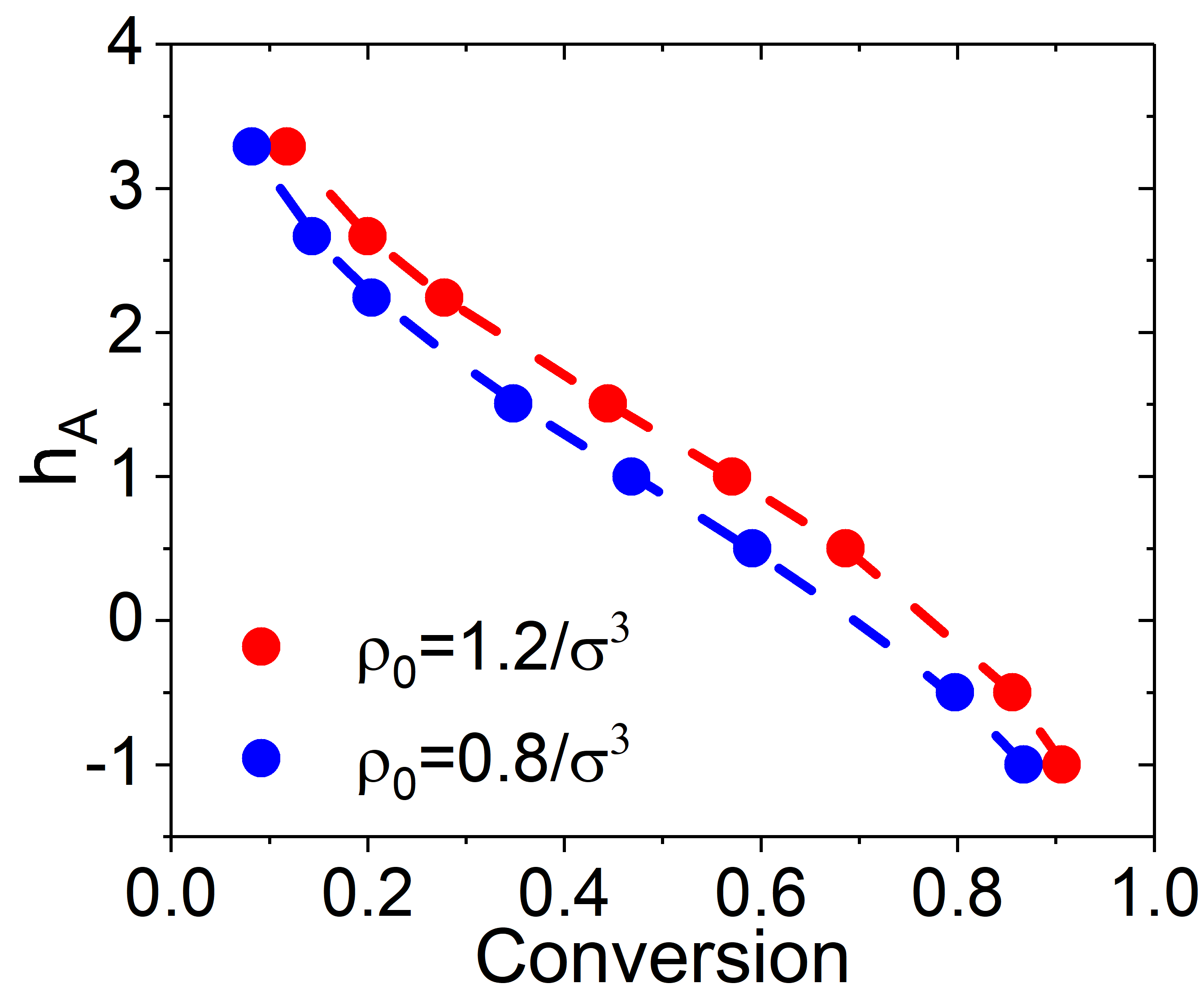}{0.98\linewidth}{$h_{A}$ is plotted against average conversion rate ($C_{e}$) at $\rho_{0}=0.8 /\sigma^{3}$ and $\rho_{0}=1.2/ \sigma^{3}$ systems, and the analytic calculation results based on eq. \ref{eq:h_above} are represented by the corresponding dashed line.}

\subsection{The Relaxation of Associative Polymer Chain}
The sampling efficiency is one of most important problems in simulation study. If the polymer chain is decorated by some solvents, the effective molecular weight will be increased. Accordingly, the chain conformation update will be dramatically slowed. Therefore, to facilitate the relaxation process caused by the increase of the effective molecular weight and branched structure, associative polymer chain configuration bias (APCCB) method is employed, of which the detail can be found in method section. The test system has one polymer chain immersed in solvent with the density being $0.8/\sigma^{3}$. $\epsilon_{\kappa}$ and $\epsilon_{\chi}$ are $0.5$ and $3.0$, respectively. In MC parameter setup, all of particles are randomly picked twice in one MC step, including one spacial movement trial and one association trial. APCCB is performed $8.72$ times on average per MC step, in which the number of internal growth trials and free end growth trials are $6.976$ and $1.744$, respectively. It is needed to mention that each APCCB trial will cost much more time than one normal hopping trial, so, only a few of APCCB trials are performed each MC step.
We take the end-to-end vector auto-correlation function (EEACF) as the indicator for chain conformation relaxation, which is a lagging indicator for chain radius of gyration correlation, and compare it between simulations with and without APCCB method. EEACF ($\phi$) at MC step interval $\Delta t$ is calculated as,\cite{huang2001dynamic}
\begin{equation}
\phi (\Delta t) = \langle {\vec{R_{e}}}(t) \cdot \vec{R_{e}}(t+\Delta t) \rangle
\end{equation}
, where $\vec{R_{e}}(t)$ is the unit vector of chain end-to-end vector at MC step $t$. Two different sets of systems are compared. One is at conversion being equal to around $0.08$, corresponding to a merely dry polymer chain. The other one is conversion at around $0.5$, corresponding to quite a wet polymer chain. The reason why we do not pick fully associated state is that if the association rate is close to $1.0$, the polymer chain conformation will be like a straight rod due to solvent steric effects, so, it is trivial to distinguish conformation change. EEACF is plotted against MC steps in figure~\ref{fig:figure_3.png} (a) for systems with and without APCCB at a conversion of $0.08$. With the incorporation of APCCB, it takes only $600$ MC steps to decrease EEACF to $0.2$ and an additional $2800$ MC steps to bring it down to approximately $0.01$. In the system without APCCB, $48000$ steps are needed to reduce EEACF to $0.2$, and a total of $168000$ MC steps to reach $0.01$. MC-step-wise, APCCB can accelerate chain relaxation by a factor of more than $50$. The actual time cost is plotted against MC steps in figure~\ref{fig:figure_3.png} (b). The APCCB system reduces the time required to reach an EEACF value of $0.2$ and $0.01$ to just $174$ seconds and $16.3$ minutes, respectively. In contrast, the system without APCCB takes $2.94$ hours and $10.43$ hours to achieve the same EEACF values, demonstrating a significant decline in efficiency. Figure~\ref{fig:figure_3.png} (c) and (d) shows the EEACF and time cost for systems with and without APCCB at a conversion of $0.5$. At higher conversion rates, the relaxation process is markedly slower. The system without APCCB requires $1164000$ MC steps to reach an EEACF value of $0.2$, and a total of $1629000$ steps to decrease EEACF to $0.01$, with time costs of $74.36$ hours and $103.9$ hours, respectively. While the system with APCCB requires only $18,000$ and $44,000$ MC steps to reach the same EEACF values, with corresponding time costs of $1.53$ hours and $3.70$ hours, respectively. The above result suggests the significant improvement of chain relaxation process by applying APCCB in the simulation.

\Figure{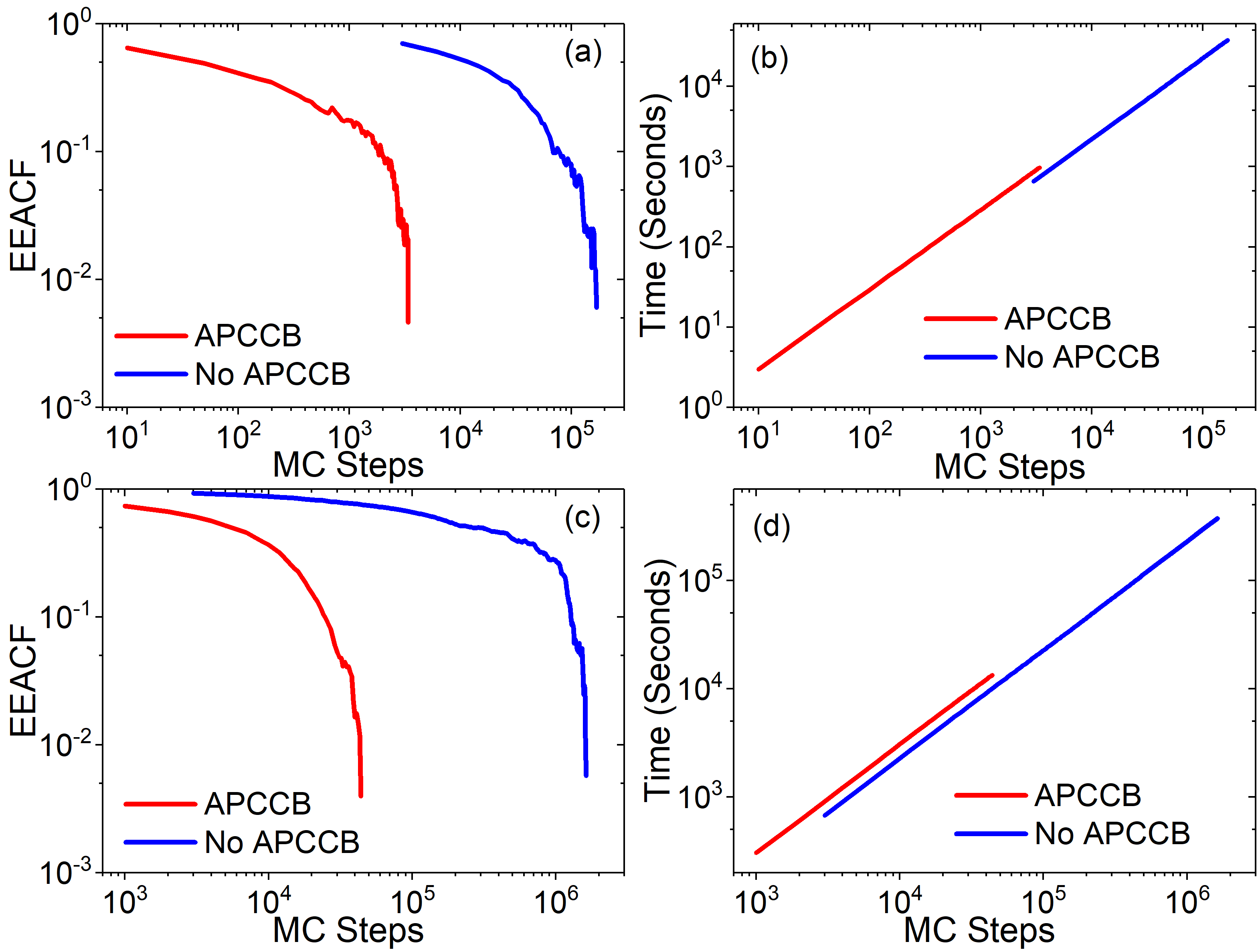}{0.98\linewidth}{The end-to-end vector auto-correlation function (EEACF) is plotted against MC steps for the average conversion of (a) $0.08$ and (c) $0.5$ systems, and the corresponding actual time cost is plotted in (b) and (d) for the conversion of $0.08$ and $0.5$ systems, respectively.}

\subsection{The Non-ideal Association System} \label{section:normal_asso}
From section \ref{sec:asso_nature}, we know that the association process in ideal system follows Bernoulli process. If we include non-bonded interactions, how will it modify the association activities? Will it still be Bernoulli process? We examine the non-ideal association system, by setting $C_{\delta h}$, cooperative association coefficient, to zero. $\epsilon_{\kappa}$ and $\epsilon_{\chi}$ are set to $0.5$ and $3.0$, respectively. We define such systems with non-bonded interactions as "real" systems. The solvent number density is $0.8 /\sigma^{3}$. 
Figure~\ref{fig:figure_4.png} shows the radius of gyration ($R_{g}^{2}$) normalized by $R_{g}^{2}$ at ideal state and average conversion as a function of $h_{A}$ at different chain lengths. $R_{g}^{2}$ is calculated based on the following equation,
\begin{equation}
R_{g}^{2}=\frac{1}{N_{P}}\sum_{k=1}^{N_{P}}|{\bf r}_{k} - {\bf r}_{\text{mean}}|^{2}
\end{equation}
, where ${\bf r}_{k}$ is the spacial coordinates of the $k$ segment, and ${\bf r}_{\text{mean}}$ is the mean position of the polymer chain. $R_{g}^{2}$ at ideal state ($R_{g0}^{2}$) is calculated by $(N_{P}-1)/6$. It can be seen that the chain conformation is coupled with the conversion rate, which has been reported widely in previous studies. \cite{okada2005cooperative,kang2016collapse,futscher2017role,zhang2019modeling,de2018molecular,gong2013modeling} The $R_{g}$ of $N_{P}=50$ system in the fully hydrated state is approximately twice that of the ideal state, and for the $N_{P}=200$ system, it is about three times larger. The attached solvents can expand chain conformation due to not only steric effects but also strong solvent-polymer repulsion, and the longer polymer chain can carry more solvents, leading to more expanded conformation. Conversion rate of all three systems almost overlap with each other at different chain lengths, suggesting that association process is only determined by $h_{A}$ value regardless of molecular weights. \par

One way to justify how the real system deviates from ideal system is to compare $h_{A}$ and $C_{e}$ relation. It is not feasible to find analytic solution for real system, so, we plot the ideal curve based on eq. \ref{eq:h_above} and shift it to fit the point. The reason why it is set highest conversion as the reference point is that the chain conformation has minimum effects at this point. By doing this, it is assumed that all of non-bonded interactions ($\epsilon_{\kappa}$ and $\epsilon_{\chi}$) come into constant term in eq. \ref{eq:h_above}, not affecting $C_{e}$ term. Figure~\ref{fig:figure_4.png} (b) compares the $h_{A}$ vs. conversion in real systems and expected ideal system, denoted by dashed line. At high and middle conversion range, corresponding to conversion larger than $0.35$ region, the analytic calculation and simulation give the similar $h_{A}$ value at the same conversion rate, indicating weak non-bonded interaction effects on association activities. At low conversion range, that is conversion lower than $0.35$ region, we can observe the deviation, that is the difference of $h_{A}$ at the same conversion (x-axis distance), and it becomes more apparent with decreasing average conversion. This difference suggests that non-bonded interactions effects not only comes into constant term in eq. \ref{eq:h_above}, but also affect first term in eq. \ref{eq:h_above}. \par

Next, we choose conversion equal to $0.35$ and $0.08$ systems at $N_{P}=200$ to plot the segments' associating probability along the chain contour index, presented in figure~\ref{fig:figure_5.png}. x-axis is the segment location along the chain contour. "1" means the head segment and "$N_{P}$" corresponds to the end segment. Because of the symmetry of the linear chain, associating probability of segments from $1$ to $N_{P}/2$ is averaged with segments from $N_{P}$ to $N_{P}/2 + 1$. There are only chain connectivity and association process in the ideal system, while the real systems have non-bonded interactions. The ideal system shows a homogeneous distribution for all three systems. The conversion rate does not depend on the segment's position, implying that all segments are indistinguishable for association process, consistent with discussion in section \ref{sec:asso_nature}. Therefore, that is perfect Bernoulli process. But for real systems, the end shows the highest associating probability, and it goes down with the segment's index moving towards the middle. The random process with different probability for each independent trial becomes Poisson binomial distribution. Moreover, the corresponding number of associated segments distribution are plotted in figure~\ref{fig:figure_6.png}. The x-axis is the total number of associated segments, and y-axis is the probability to observe the corresponding $x$. The point of ideal systems lies exactly on the binomial distribution. However, the peak of real systems is slightly lower than the ideal system due to inhomogeneous associating probability, that is Poisson binomial distribution. 
This effect exists across the entire range of conversion systems regardless of conversion rate, as polymer segments will always occupy solvent positions. Two ends have more contact area with solvents, while mid-segments are wrapped inside the polymer chain. However, the inhomogeneous associating probability plays a more significant role in low-conversion systems due to more collapsed conformation, as indicated by the large deviation in figure~\ref{fig:figure_4.png} (b). \par

By summarizing all of the above observations for real systems, attached solvents can expand the chain conformation not only due to steric effects but also solvents immiscibility, and the association process does not show any dependence on molecular weight. Compared with ideal system, the excluded volume and solvents immiscibility effects can change the segment association probability at various positions by changing the local association candidates density, and in further turn binomial distribution to Poisson binomial distribution. Moreover, non-bonded interactions do not introduce correlation to the association process, as the distribution shape remains unchanged qualitatively. \par

\Figure{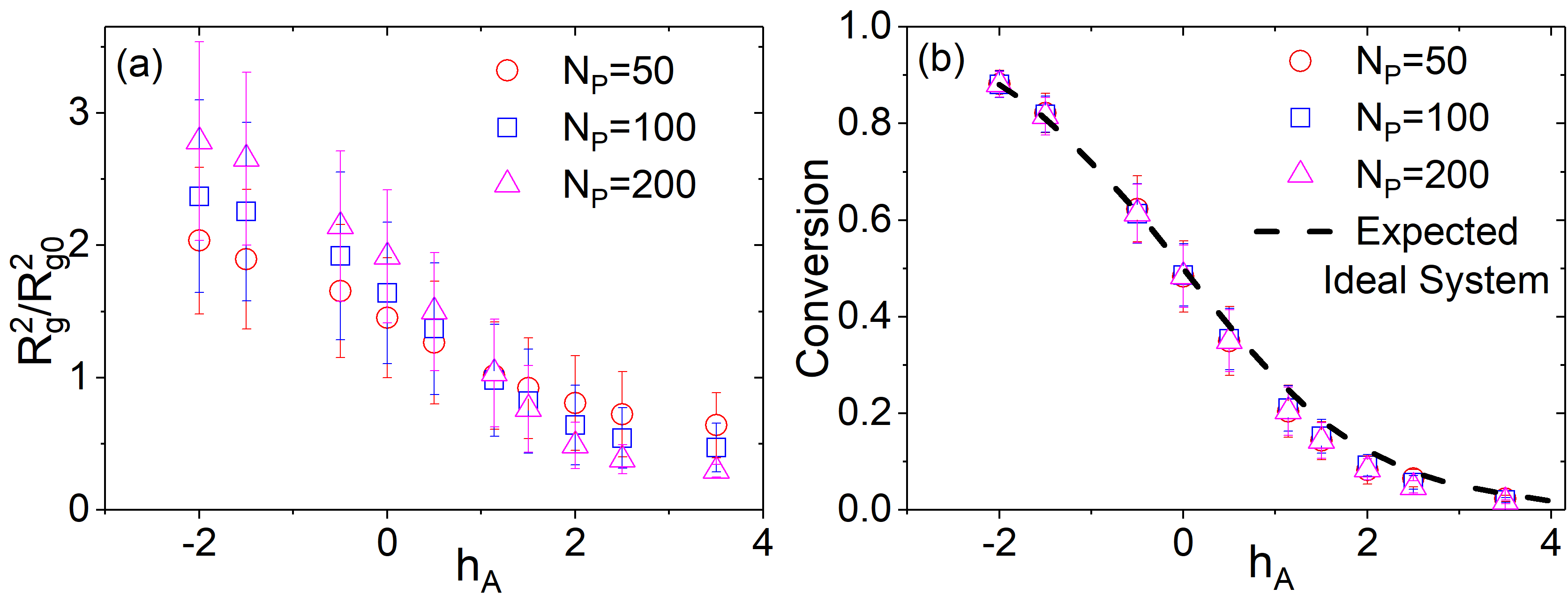}{0.98\linewidth}{(a) Radius of gyration ($R_{g}^{2}$) normalized by the ideal state $R_{g0}^{2}$ is plotted against the association energy barrier ($h_{A}$) with $C_{\delta h}=0$ at different chain lengths. (b) The average conversion is plotted against association energy barrier ($h_{A}$) with $C_{\delta h}=0$ at different chain lengths.}
\Figure{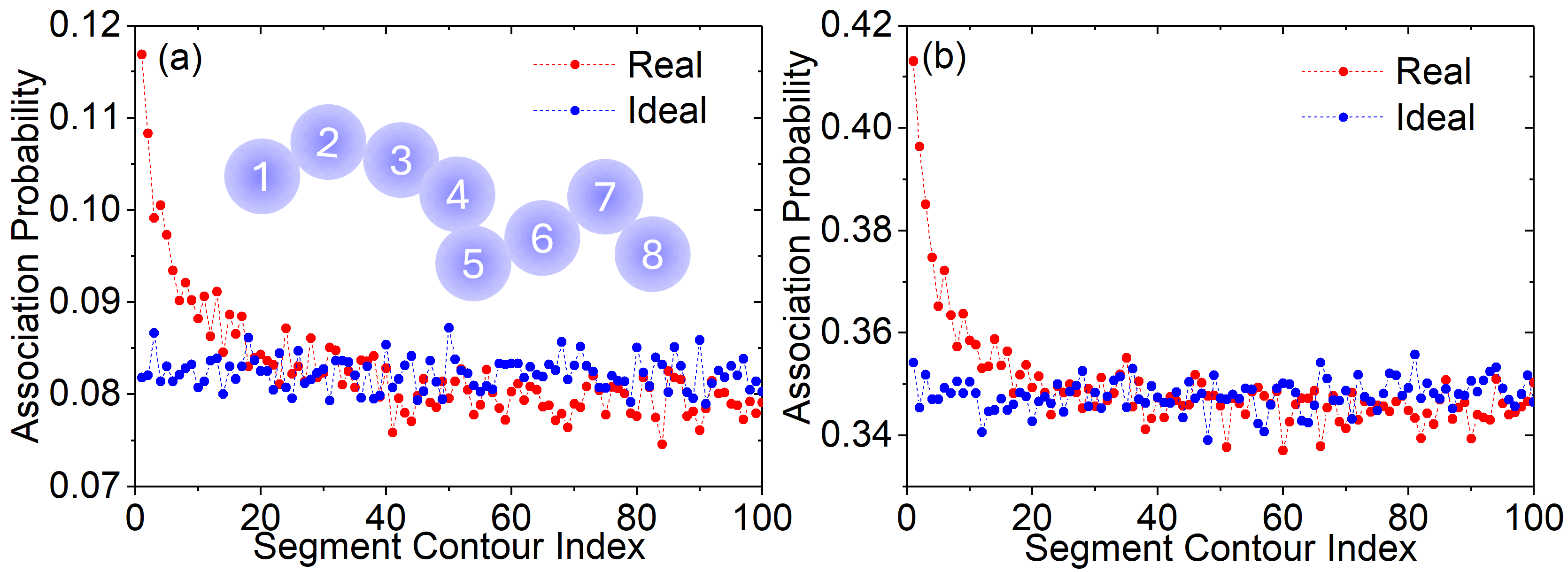}{0.98\linewidth}{The associating probability of each segment along the chain contour is plotted for the average conversion of (a) $0.08$ and (b) $0.35$ systems with $N_{P}=200$. The inset plot shows an example of the chain contour index.}
\Figure{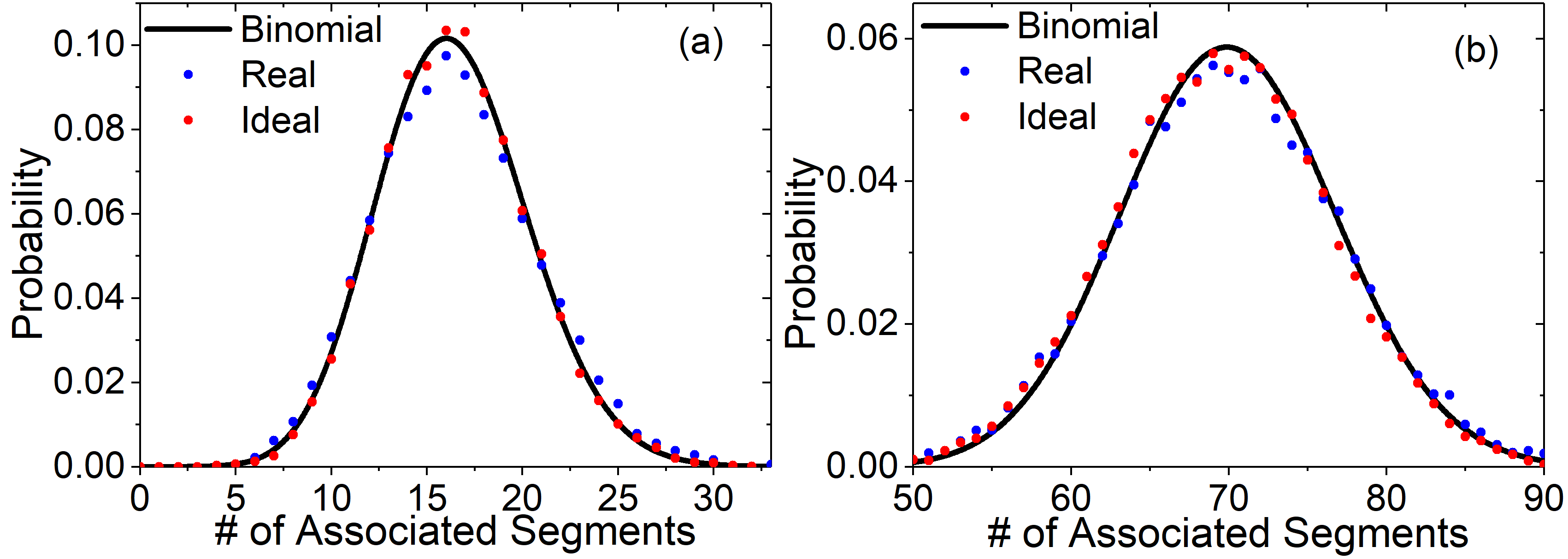}{0.98\linewidth}{The probability distribution of the number of associated segments at the average conversion of (a) $0.08$ and (b) $0.35$ with $N_{P}=200$.}

\subsection{The Cooperative Association System}
Next, we extend the normal association process to cooperative association systems. To improve sampling efficiency, replica exchange method is used, the detail of which is included in appendix. In cooperative association study, $h_{A}$ is set to be a constant, equal to $6.5$, which means that the base association barrier is very high. The average conversion rate in $h_{A}=6.5$, $C_{\delta h}=0$ system is $0.001$, and $C_{\delta h}$ is the variable. So, the association is basically induced by the cooperativity, that is $C_{\delta h}$. Figure~\ref{fig:figure_7.png} shows the radius of gyration normalized by ideal state $R_{g0}^{2}$ and conversion as a function of $C_{\delta h}$. There are two common features between non-cooperative system and cooperative association system. The chain conformation is coupled with the association rate, and the conformation change becomes more evident with the increase of the chain length. In non-cooperative system, $R_{g}^{2}$ curves of three different chain lengths cross at one point, indicating that the molecular weight has a weak effect on the transition point. But in cooperative system, an evident shift of the conversion and $R_{g}^{2}$ curve with the chain length can be observed. So, the introduction of the cooperative association will significantly affect the molecular weight effect on the transition point. Based on experimental results, the transition point of some polymers, like PNIPAM (Poly(N-isopropylacrylamide)), has the weak dependence on molecular weight \cite{furyk2006effects}. But for some other systems, like $\text{PVPip}$ (poly(N-vinylpiperidone)), the transition point strongly depends on the molecular weight \cite{ieong2011missing}. In light of our observation, the strength of cooperativity decides the dependence of transition point on molecular weight.\par

It is well-acknowledged that the polymer conformational transition accompanied with dehydration or hydration is first order transition, as hysteresis upon cooling and heating can be observed.\cite{zhang2017thermoresponsive, wu1998globule, ray2005effect, hirano2006dual, mukherji2017reply, de2018molecular}
Moreover, coil and globule state may coexist at the transition point during first order transition. In our simulation results, it can be seen that the error bar in figure~\ref{fig:figure_7.png} is much larger than it in non-cooperative systems, suggesting the large fluctuation. To take a closer look, conversion distribution and radius of gyration distribution at conversion of $0.5$ systems are plotted in figure~\ref{fig:figure_8.png}, where the steepest slope can be detected. In conversion distribution plot, all three systems exhibit a wide and deep minimum at the midpoint, with two peaks at ends of the curve, indicating the coexistence of two associating states. As previously shown, the association rate strongly couples with the chain conformation, so, corresponding double-peak behaviors should also be observed in radius of gyration distribution plot. However, the $R_{g}^{2}$ distribution plot suggests the qualitatively different behaviors. The polymer configuration state cannot be well distinguished when the chain length is not long enough, as only one broad peak can be observed for $N_{P}=50$ system. With the increase of the molecular weight, the minimum point between two peaks becomes deeper, and the coexistence of coil and collapsed states becomes more apparent. A previous theoretical study reports a similar observation, noting that the double-peak behavior is more pronounced with increasing chain length, though in their study, the change in the order of the transition is attributed to the force constant of non-bonded contacts.\cite{maffi2012first} \par

The umbrella sampling method is used to calculate the potential of mean force at the transition point to verify the stability at two states. The details about umbrella sampling is shown in appendix. $R_{g}^{2}$ is chosen as the reaction coordinates, and a range of $R_{g}^{2}$ is scanned. It is known that when the restraint force is too weak, the potential of mean force (PMF) curve fails to accurately reflect the correct probability distribution, resulting in a lack of observable features. Conversely, if the restraint force is too strong, the system is constrained at a single state, losing thermodynamic fluctuations. Therefore, the set of restraint simulation is running with different restraint potential coefficient, $C_{um}$, and the most proper one is chosen. The result is shown in figure~\ref{fig:figure_9.png} (a). The PMF curve only exhibits single peak in chain length equal to $50$ system. In chain length equal to $100$ system, one clear peak is shown at $R_{g}^{2}/R_{g0}^{2}=0.5$, and the other peak can be roughly observed at $R_{g}^{2}/R_{g0}^{2}=2$. The double peak of PMF curve can be clearly remarked at chain length equal to $200$ system. Only when the chain length is long enough, the coexistence of two states can be noticed, which is also suggested by $R_{g}^{2}$ distribution plot. The position of the peak observed in PMF plot is basically consistent with the $R_{g}^{2}$ distribution plot. The coexistence arises from the competition between large association energy barrier and strong cooperative strength as it is shown in schematic plot, that is  figure~\ref{fig:figure_9.png} (b). The fully dehydrated and fully hydrated states each have their own advantages over the other. While the segment is reluctant to associate with solvents due to a high energy barrier, there is always a low probability that a single segment becomes hydrated. Once this occurs, the entire chain quickly becomes fully hydrated due to strong cooperative interactions. Therefore, it can be conjectured that cooperative association can lead to the coexistence of coil and globule state at the transition point.

\Figure{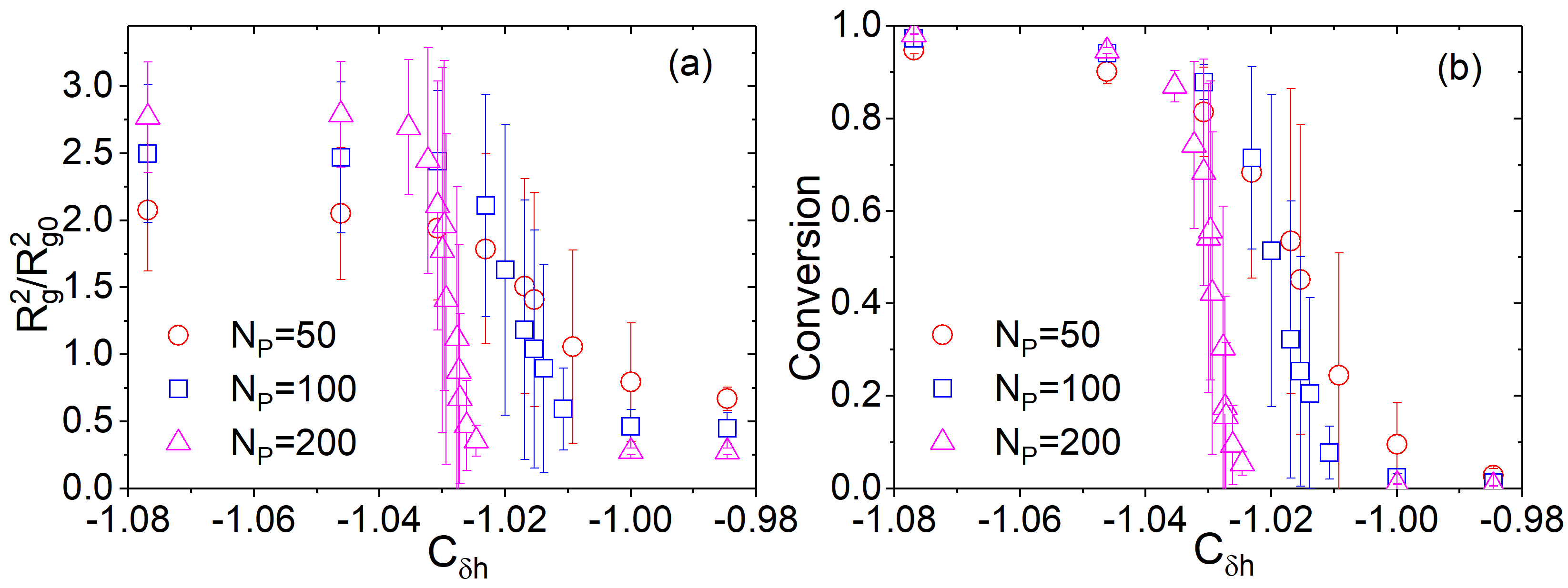}{0.98\linewidth}{(a) Radius of gyration ($R_{g}^{2}$) normalized by the ideal state $R_{g0}^{2}$ is plotted against the cooperative coefficient ($C_{\delta h}$) with $h_{A}=6.5$ at different chain lengths. (b) The average conversion rate is plotted against the cooperative coefficient ($C_{\delta h}$) with $h_{A}=6.5$ at different chain lengths.}
\Figure{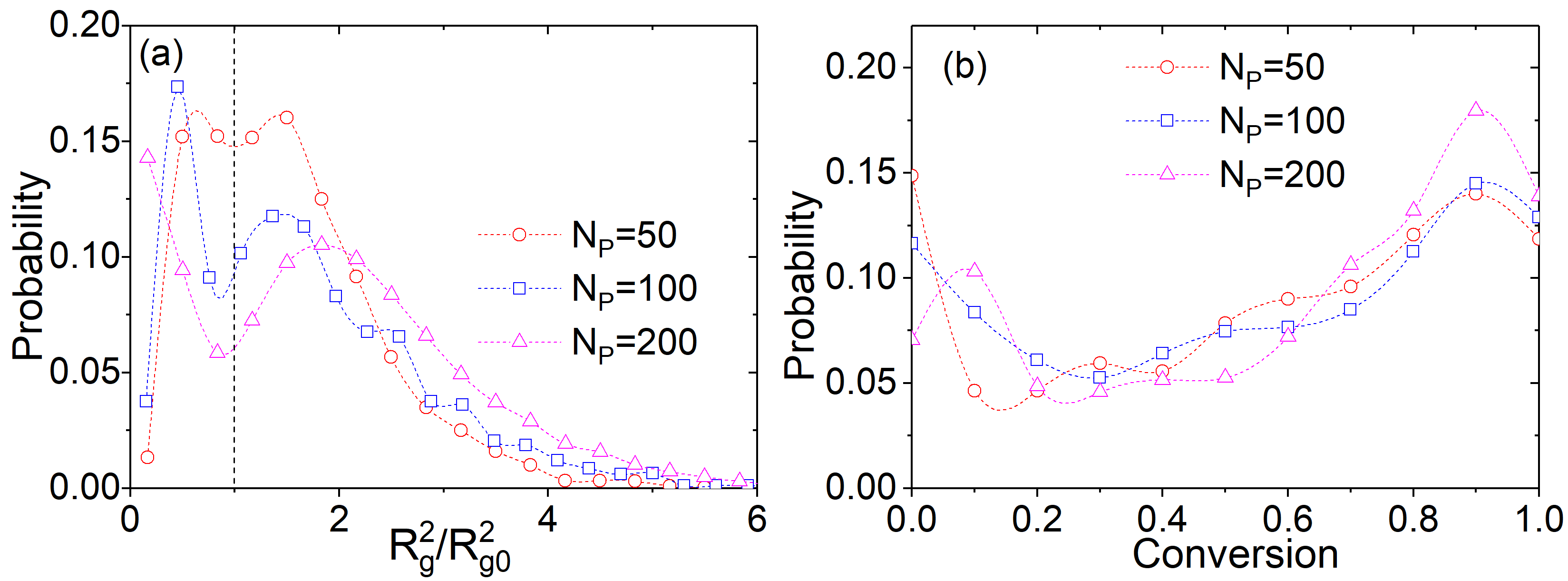}{0.98\linewidth}{(a) Distribution of radius of gyration normalized by ideal state $R_{g0}^{2}$ in cooperative association system with average conversion equal to $0.5$ at different chain lengths. (b) Conversion distribution in cooperative association system with conversion equal to $0.5$ at different chain lengths.}
\Figure{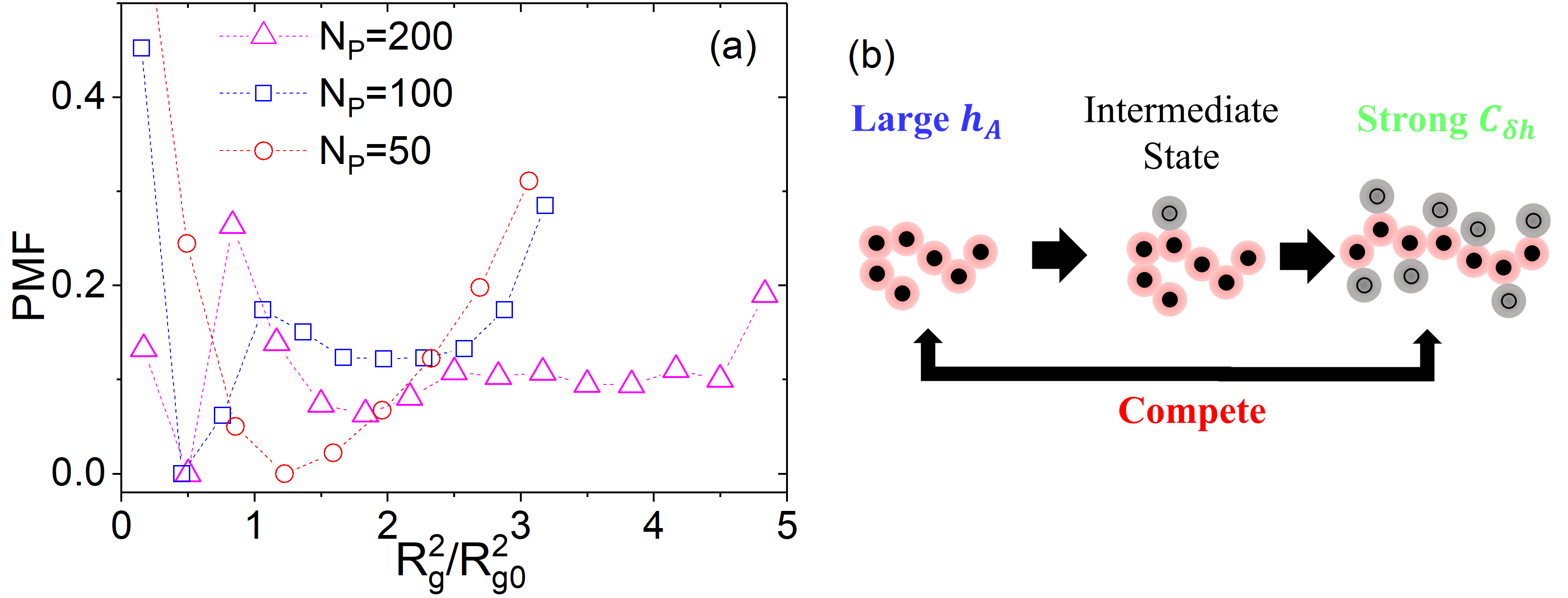}{0.98\linewidth}{(a) Free energy along radius of gyration (potential of mean force) calculated by umbrella sampling method plotted against $R_{g}^{2}/R_{g0}^{2}$ at conversion equal to $0.5$ with $N_{P}=50$ ($C_{um}=0.02$), $N_{P}=100$ ($C_{um}=0.02$) and $N_{P}=200$ ($C_{um}=0.01$). (b) The schematic representation of competition process between large association energy barrier and strong cooperative association strength.}

\subsection{Mathematical Solution Discussion For Association Behaviors}
It is not feasible to find out all possible associative patterns, therefore, one-mode approximation is commonly used to study association problems, in which only the most likely sequence pattern is considered \cite{okada2005cooperative}. But we can move further based on the method about calculating the sequence distribution \cite{schilling1990longest}. There are totally $2^{N_{P}}$ possible associating patterns for one polymer chain. The longest sequence length in each pattern can be found by using the recursion algorithm discussed in the cited paper, and other shorter sequences in the pattern will not be considered \cite{schilling1990longest}. The equation for the number of sequences less than $m$ with totally $k$ segments associated can be written as \cite{schilling1990longest}, 
\begin{equation}
C_{N_{P}}^{k}(m)= \begin{cases}
\sum_{j=0}^{m} C_{N_{P}-j-1}^{k-1}(m), \quad \text{for} N_{P}>k>m \\
{\binom{N_{P}}{k}}, \quad \text{for} k \leq m \\
0, \quad \text{for} m<k=N_{P} .
\end{cases}
\end{equation}
So, the number of a certain sequence length can be written as, 
\begin{equation}
P(k,m)=	C_{N_{P}}^{k}(m) - C_{N_{P}}^{k}(m-1) . 
\end{equation}
Next, the energy weight can be added to the equation as following, 
\begin{equation}
T(k,m)=P(k,m)\exp{(-h \cdot C_{\delta h} h_{A})} \exp{(-k\cdot h_{A})}
\end{equation}
where $h=0$, if $m \leq 1$, and $h=m-1$, for other situations. The normalized distribution can be written as,
\begin{equation}
\tilde{T}(k,m)= \frac{T(k,m)}{\sum_{k}\sum_{m}T(k,m)}  .
\end{equation}
The analytical solution for the conversion rate distribution with both $h_{A}$ and $C_{\delta h}$ included can be express as $D(k)=\sum_{m}\tilde{T}(k,m)$. Next, two cases — non-cooperative association and cooperative association systems, both including non-bonded interactions — are tested for $N_{P}=30$ systems, as the recursion process and factorial make the numerical calculation very difficult for longer chains. The overall conversion rate is plotted in figure~\ref{fig:figure_10.png}. It can be seen that conversion calculated from MC simulation couples well with the theory prediction in both non-cooperative system and cooperative system. In cooperative association system, a small extent of deviation can be observed, which is caused by the longest sequence approximation. 


Next, the distribution in MC simulation and analytical calculation is plotted in figure~\ref{fig:figure_11.png} for $h_{A}=0, C_{\delta h} = 0$ system (non-cooperative association) and $h_{A}=6.5, C_{\delta h}=-6.7$ system (cooperative association). Certainly, the distribution for non-cooperative association system is close to binomial distribution. Additionally, the analytical results align well with the Monte Carlo simulations for the cooperative association system, demonstrating double-peak behavior with similar peak positions. The consistency between analytic model and MC simulation indicates that the driving force for the coexistence of two-states is the competition between $h_{A}$ and $C_{\delta h}$. The association entropy (statistical distribution weight) drives the single peak behavior, which is binomial-like distribution. However, when enthalpy bias is added to the association activities, the coexistence of two states can be observed.

\par

\Figure{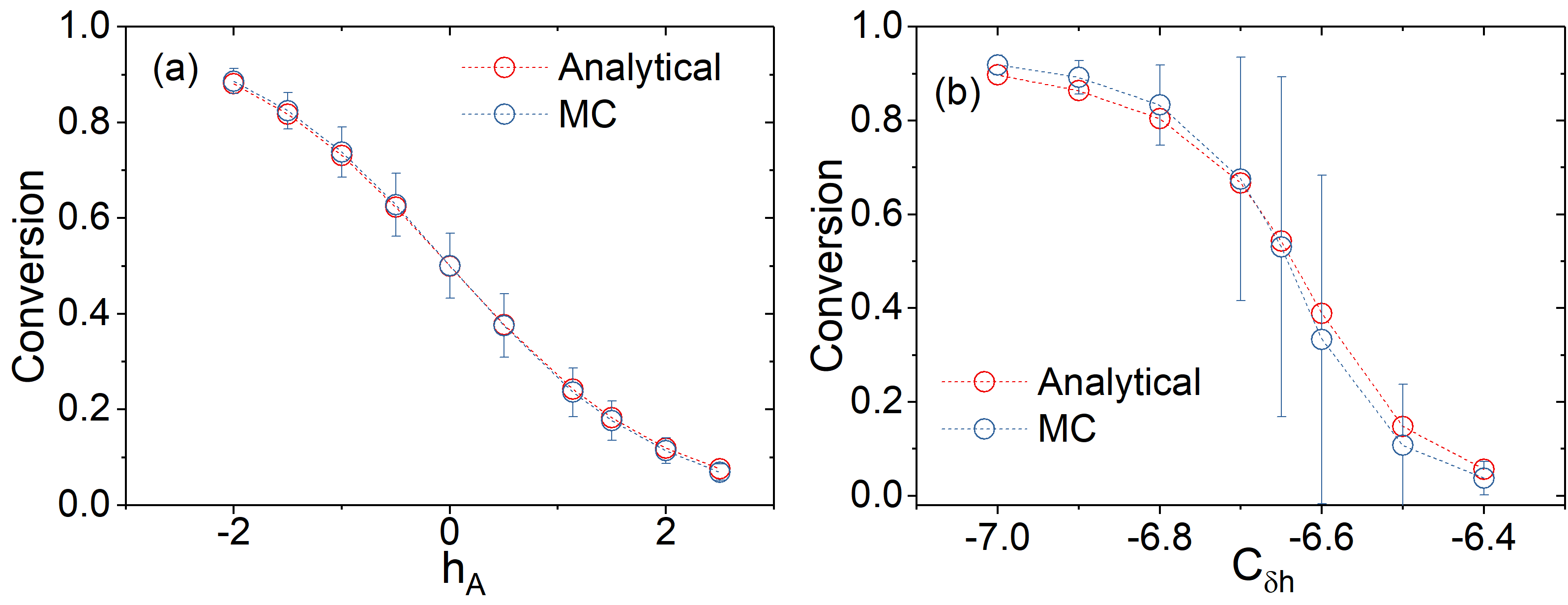}{0.98\linewidth}{Conversion rate calculated from MC simulation and analytical model plotted against the (a) $h_{A}$ with $C_{\delta h}=0$ and (b) $C_{\delta h}$ with $h_{A}=6.5$ for chain length equal to $30$ system. } 

\Figure{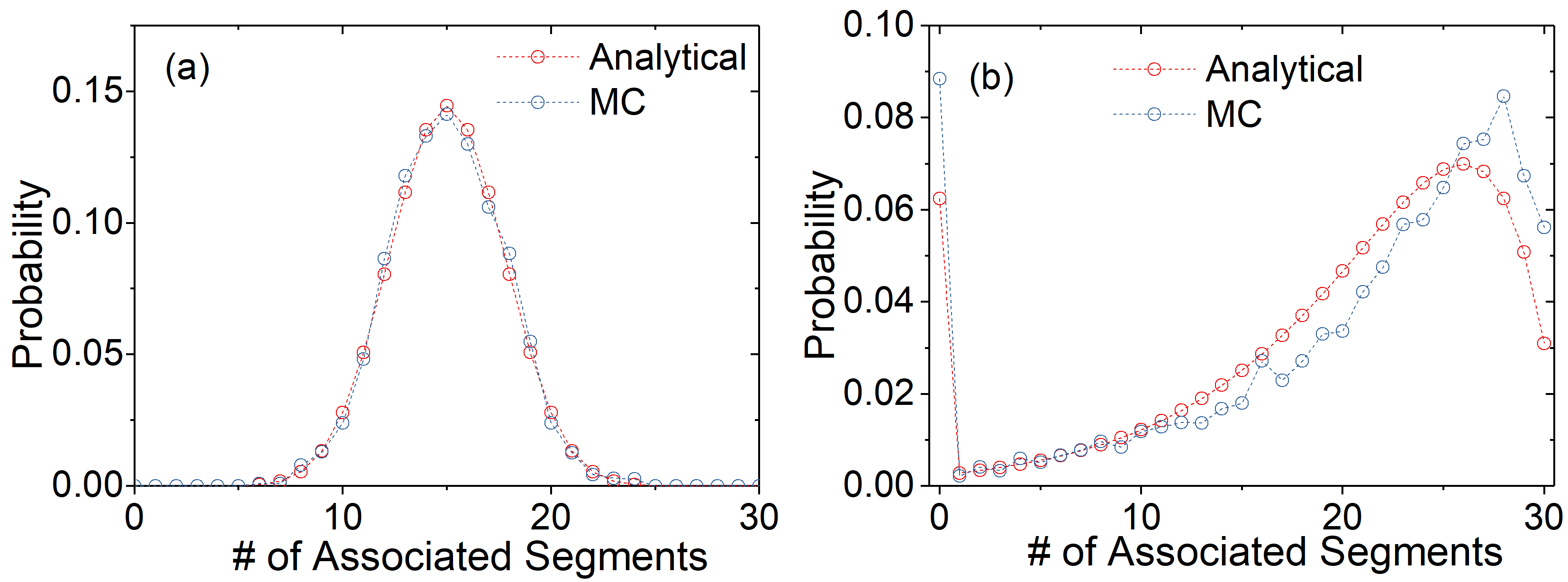}{0.98\linewidth}{The number of associated segments distribution calculated from MC simulation and analytical model plotted for (a) $h_{A}=0, C_{\delta h} = 0$ system (non-cooperative association) and (b) $h_{A}=6.5, C_{\delta h}=-6.7$ system (cooperative association).} 

\section{Conclusion}
We propose a reaction-controlled model to investigate polymer associating behaviors in solvents. Correspondingly, APCCB method is developed to help the chain relaxation, proving that it can significantly improve sampling efficiency. The pure association process in ideal systems follows Bernoulli process, confirmed by analytic calculation and simulation results. And the association activities in our model are demonstrated to intrinsically obey the principles of thermodynamics. When the non-bonded interactions is present, the chain conformation transition is investigated with respect to $h_{A}$, induced by the change of the number of attached solvents. It turns out that excluded volume and solvents immiscibility effects can change the associating probability at different locations along the chain contour, leading to Poisson binomial distribution instead of binomial distribution for ideal systems. However, non-bonded interactions do not bring correlations to the association process. We extend the study to cooperative association systems. It is found that the cooperative association can lead to the strong dependence of transition point on molecular weight and the sharp transition process, while the transition point is independent of molecular weight in non-cooperative systems. In further, the coexistence of coil and globule states can be observed at the transition point when the chain is long enough, the stability of which is verified by free energy calculation along the radius of gyration. At last, the mathematical model discussion confirms that the association entropy (combinatorial probability) only gives the single peak behaviors, while the addition of enthalpy bias can lead to double peak behaviors. This study provides a through and insightful analysis for association process. We believe this will contribute to the development of future association models and the rational design of association-related materials.

\section{Appendix}
\subsection{Replica Exchange Method}
To help cross the energy barrier in cooperative association systems, replica exchange method is used, which is based on the scheme developed by previous works. \cite{swendsen1986replica,sugita1999replica} the association energy parameters ($h_{A}$ and $\delta h$) are exchanged in this study, and the corresponding acceptance criterion is derived as following. 
\begin{equation}
\begin{split}
& P({\bf i},x_{i})P({\bf j},x_{j})T[({\bf i},x_{i}),({\bf j},x_{j}) \rightarrow ({\bf j},x_{i}),({\bf i},x_{j})]acc[({\bf i},x_{i}),({\bf j},x_{j}) \rightarrow ({\bf j},x_{i}),({\bf i},x_{j})]  \\
& = P({\bf i},x_{j})P({\bf j},x_{i})T[({\bf i},x_{j}),({\bf j},x_{i}) \rightarrow ({\bf j},x_{j}),({\bf i},x_{i})]acc[({\bf i},x_{j}),({\bf j},x_{i}) \rightarrow ({\bf j},x_{j}),({\bf i},x_{i})] \\
\end{split}
\end{equation}
where $P({\bf i},x_{i})$ is the probability to observe the ${\bf i}$ configuration at association parameter $x_{i}$, and $T$ is the proposed transiting probability. The random choices of replicas ensures the symmetry of $T$. So, the acceptance criterion can be written as, 
\begin{equation}
\label{eq:re_acce}
acc[({\bf i},x_{i}),({\bf j},x_{j}) \rightarrow ({\bf j},x_{i}),({\bf i},x_{j})] = \min(1, \frac{P({\bf i},x_{j})P({\bf j},x_{i})}{P({\bf i},x_{i})P({\bf j},x_{j})})
\end{equation}
The Hamiltonian of the system can be described by three terms,
\begin{equation}
\mathcal{H}=\mathcal{H}^{b}+\mathcal{H}^{nb}+\mathcal{H}^{a}
\end{equation}
where $\mathcal{H}^{b}$ is the total bonded energy, $\mathcal{H}^{nb}$ is the total non-bonded energy and $\mathcal{H}^{a}$ is the total association energy. And $P({\bf i},x_{i})$ can be written as,
\begin{equation}
P({\bf i},x_{i}) = \frac{\exp[-(\mathcal{H}^{b}({\bf i},x_{i})+\mathcal{H}^{nb}({\bf i},x_{i})+\mathcal{H}^{a}({\bf i},x_{i}))]}{\mathcal{Z}}
\end{equation}
where $\mathcal{H}^{b(nb \, \text{or} \, a)}({\bf i},x_{i})$ represents the corresponding energy (bonded, non-bonded or association energy) in ${\bf i}$ configuration with parameter $x_{i}$. According to Eq. \ref{eq:re_acce}, the acceptance criterion can be expressed as,
\begin{equation}
\begin{split}
& acc[({\bf i},x_{i}),({\bf j},x_{j}) \rightarrow ({\bf j},x_{i}),({\bf i},x_{j})]  =  
 \exp \{ \mathcal{H}^{b}({\bf i},x_{i})+\mathcal{H}^{nb}({\bf i},x_{i})+\mathcal{H}^{a}({\bf i},x_{i})+\mathcal{H}^{b}({\bf j},x_{j})+ \\
& \mathcal{H}^{nb}({\bf j},x_{j})+\mathcal{H}^{a}({\bf j},x_{j}) 
 - \mathcal{H}^{b}({\bf i},x_{j})-\mathcal{H}^{nb}({\bf i},x_{j})-\mathcal{H}^{a}({\bf i},x_{j})-\mathcal{H}^{b}({\bf j},x_{i})-\mathcal{H}^{nb}({\bf j},x_{i})-\mathcal{H}^{a}({\bf j},x_{i}) \} 
\end{split}
\end{equation}
Because we are only exchanging $h_{A}$ and $\delta h$, the total covalent bond energy and non-bonded energy are the same at the same configuration. Accordingly, the equation can be simplified to,
\begin{equation}
acc[({\bf i},x_{i}),({\bf j},x_{j}) \rightarrow ({\bf j},x_{i}),({\bf i},x_{j})] = \exp \{ \mathcal{H}^{a}({\bf i},x_{i}) + \mathcal{H}^{a}({\bf j},x_{j}) - \mathcal{H}^{a}({\bf i},x_{j}) - \mathcal{H}^{a}({\bf j},x_{i}) \}
\end{equation}
According to Eq. \ref{eq: asso_total} and Eq. \ref{eq: asso_pair}, it can be observed that the first term, that is the bond length depended term, is equal at the same configuration. So, the above equation can be further simplified to,
\begin{equation}
\begin{split}
& acc[({\bf i},x_{i}),({\bf j},x_{j}) \rightarrow ({\bf j},x_{i}),({\bf i},x_{j})] = \exp \{ - \sum_{l=1}^{N_{asso}({\bf i})}(h_{A}({\bf j})+\delta h({\bf j}) \cdot s_{P,l}({\bf i})) \\
& - \sum_{l=1}^{N_{asso}({\bf j})}(h_{A}({\bf i})+\delta h({\bf i}) \cdot s_{P,l}({\bf j})) + \sum_{l=1}^{N_{asso}({\bf i})}(h_{A}({\bf i})+\delta h({\bf i}) \cdot s_{P,l}({\bf i})) \\
& + \sum_{l=1}^{N_{asso}({\bf j})}(h_{A}({\bf j})+\delta h({\bf j}) \cdot s_{P,l}({\bf j})) \}
\end{split}
\end{equation}
where $N_{asso}({\bf i})$ is the number of association bonds at ${\bf i}$ configuration, $h_{A,i}$ and $\Delta h_{i}$ are association parameters at condition $x_{i}$, and $s_{P,l}({\bf i})$ is the association states of $l^{\mathrm{th}}$ associated polymer segment at ${\bf i}$ configuration.

\subsection{Umbrella Sampling} 
Umbrella sampling method is applied to calculate the potential of mean force at different states. Radius of gyration of the polymer chain is chosen as the reaction coordinate. So, the total Hamiltonian of the system becomes the following equation with the addition of restraint potential,
\begin{equation}
\mathcal{H}=\mathcal{H}^{b}+\mathcal{H}^{nb}+\mathcal{H}^{a}+v_{W,k}(R_{g}^{2})
\end{equation}
and $v_{W,k}(R_{g}^{2})$ is equal to,
\begin{equation}
v_{W,k}(R_{g}^{2}) = \frac{1}{2}C_{um}(R_{g,k}^{2}-R_{g}^{2})^{2}
\end{equation}
where $k$ is the index of a set of restraint simulations from $1$ to $n$, ${R_{g,k}^{2}}$ is the reference radius of gyration in the $k^{\mathrm{th}}$ simulation, and $C_{um}$ is restraint potential strength. The way to calculate the potential of mean force follows the description in "section 9.2.3", "Computer Simulation of Liquids, 2nd edition". \cite{allen2017computer} For a particular $k$, the biased probability distribution is,
\begin{equation}
\rho_{W,k}(R_{g}^{2})=\exp{(A_{k})} \int d{\bf r} \exp{(-\mathcal{H} - v_{W,k}(R_{g}^{2}({\bf r})))} \delta(R_{g}^{2}({\bf r})-R_{g,k}^{2})
\end{equation}
where $A_{k}$ represents the free energy of the system with restraint potential implemented. $\rho_{W,k}$ can be calculated directly in the simulation,
\begin{equation}
\rho_{W,k}(R_{g}^{2}) = \frac{\left\langle S(R_{g}^{2}({\bf r}) - R_{g}^{2}, \Delta R_{g}^{2}) \right\rangle}{\Delta R_{g}^{2} M_{k}} , S =
\left\{
	\begin{array}{ll}
  		1 &  |R_{g}^{2}({\bf r}) - R_{g}^{2}| < \frac{1}{2} \Delta R_{g}^{2}  \\
  		0 &  \textnormal{otherwise}
 	\end{array} 
\right.
\end{equation}
where $S$ sorts the radius of gyration, $R_{g}^{2}({\bf r})$, into bins of width $\Delta R_{g}^{2}$ around $R_{g}^{2}$, and $M_{k}$ is the total number of histogram entries. By using weighted histogram analysis method and Lagrange multipliers, the unbiased distribution of radius of gyration can be found,\cite{allen2017computer,kumar1992weighted} 
\begin{equation}
\rho(R_{g}^{2})=\frac{\sum_{k=1}^{n} M_{k}\rho_{W,k}(R_{g}^{2})}{\sum_{k=1}^{n}M_{k} \exp{(A_{k}-A_{0})} \exp{(-v_{W,k}(R_{g}^{2}))} }
\end{equation}
where $A_{0}$ is the free energy of the unbiased system and,
\begin{equation}
\exp{(A_{k}-A_{0})}=\int d R_{g}^{2} \rho(R_{g}^{2}) \exp{(-v_{W,k}(R_{g}^{2}))}
\end{equation}
The above equations with two unknown variables $\exp{(A_{k}-A_{0})}$ and $\rho(R_{g}^{2})$ are solved by using Broyden mixing method in our study.

\newpage
\bibliographystyle{unsrt}
\bibliography{zz_citation}

\begin{thebibliography}{10}

\bibitem{song2020high}
Pingan Song and Hao Wang.
\newblock High-performance polymeric materials through hydrogen-bond
  cross-linking.
\newblock {\em Advanced Materials}, 32(18):1901244, 2020.

\bibitem{neo2016conjugated}
Wei~Teng Neo, Qun Ye, Soo-Jin Chua, and Jianwei Xu.
\newblock Conjugated polymer-based electrochromics: materials, device
  fabrication and application prospects.
\newblock {\em Journal of Materials Chemistry C}, 4(31):7364--7376, 2016.

\bibitem{he2015nanomedicine}
Chunbai He, Demin Liu, and Wenbin Lin.
\newblock Nanomedicine applications of hybrid nanomaterials built from
  metal--ligand coordination bonds: nanoscale metal--organic frameworks and
  nanoscale coordination polymers.
\newblock {\em Chemical reviews}, 115(19):11079--11108, 2015.

\bibitem{potaufeux2020comprehensive}
Jean-Emile Potaufeux, J{\'e}r{\'e}my Odent, Delphine Notta-Cuvier, Franck
  Lauro, and Jean-Marie Raquez.
\newblock A comprehensive review of the structures and properties of ionic
  polymeric materials.
\newblock {\em Polymer Chemistry}, 11(37):5914--5936, 2020.

\bibitem{zhang2015polymers}
Qilu Zhang and Richard Hoogenboom.
\newblock Polymers with upper critical solution temperature behavior in
  alcohol/water solvent mixtures.
\newblock {\em Progress in Polymer Science}, 48:122--142, 2015.

\bibitem{zhang2017thermoresponsive}
Qilu Zhang, Christine Weber, Ulrich~S Schubert, and Richard Hoogenboom.
\newblock Thermoresponsive polymers with lower critical solution temperature:
  from fundamental aspects and measuring techniques to recommended turbidimetry
  conditions.
\newblock {\em Materials Horizons}, 4(2):109--116, 2017.

\bibitem{Zhang2025Unraveling}
Xiangyu Zhang, Jing Zong, and Dong Meng.
\newblock Unraveling the mechanism for polymer cosolvency in binary mixed
  solvents, 2025.
\newblock Unpublished manuscript.

\bibitem{yan2016kinetic}
Yun Yan, Jianbin Huang, and Ben~Zhong Tang.
\newblock Kinetic trapping--a strategy for directing the self-assembly of
  unique functional nanostructures.
\newblock {\em Chemical Communications}, 52(80):11870--11884, 2016.

\bibitem{evans2018self}
Katherine Evans and Ting Xu.
\newblock Self-assembly of supramolecular thin films: Role of small molecule
  and solvent vapor annealing.
\newblock {\em Macromolecules}, 52(2):639--648, 2018.

\bibitem{hoy2009thermoreversible}
Robert~S Hoy and Glenn~H Fredrickson.
\newblock Thermoreversible associating polymer networks. i. interplay of
  thermodynamics, chemical kinetics, and polymer physics.
\newblock {\em The Journal of chemical physics}, 131(22), 2009.

\bibitem{wang2008reversible}
Shihu Wang, Chun-Chung Chen, and Elena~E Dormidontova.
\newblock Reversible association and network formation in 3: 1 ligand--metal
  polymer solutions.
\newblock {\em Soft Matter}, 4(10):2039--2053, 2008.

\bibitem{sciortino2017three}
Francesco Sciortino.
\newblock Three-body potential for simulating bond swaps in molecular dynamics.
\newblock {\em The European Physical Journal E}, 40:1--4, 2017.

\bibitem{rovigatti2018self}
Lorenzo Rovigatti, Giovanni Nava, Tommaso Bellini, and Francesco Sciortino.
\newblock Self-dynamics and collective swap-driven dynamics in a particle model
  for vitrimers.
\newblock {\em Macromolecules}, 51(3):1232--1241, 2018.

\bibitem{daoulas2009phase}
Kostas~Ch Daoulas, Anna Cavallo, Roy Shenhar, and Marcus M{\"u}ller.
\newblock Phase behaviour of quasi-block copolymers: A dft-based monte-carlo
  study.
\newblock {\em Soft Matter}, 5(22):4499--4509, 2009.

\bibitem{wittmer1998dynamical}
JP~Wittmer, A~Milchev, and ME~Cates.
\newblock Dynamical monte carlo study of equilibrium polymers: Static
  properties.
\newblock {\em The Journal of chemical physics}, 109(2):834--845, 1998.

\bibitem{milchev2000dynamical}
A~Milchev, JP~Wittmer, and DP~Landau.
\newblock Dynamical monte carlo study of equilibrium polymers: Effects of high
  density and ring formation.
\newblock {\em Physical Review E}, 61(3):2959, 2000.

\bibitem{chen2004ring}
Chun-Chung Chen and Elena~E Dormidontova.
\newblock Ring- chain equilibrium in reversibly associated polymer solutions:
  Monte carlo simulations.
\newblock {\em Macromolecules}, 37(10):3905--3917, 2004.

\bibitem{chen2006monte}
Chun-Chung Chen and Elena~E Dormidontova.
\newblock Monte carlo simulations of end-adsorption of head-to-tail reversibly
  associated polymers.
\newblock {\em Macromolecules}, 39(26):9528--9538, 2006.

\bibitem{okada2005cooperative}
Yukinori Okada and Fumihiko Tanaka.
\newblock Cooperative hydration, chain collapse, and flat lcst behavior in
  aqueous poly (n-isopropylacrylamide) solutions.
\newblock {\em Macromolecules}, 38(10):4465--4471, 2005.

\bibitem{koizumi2004concentration}
Satoshi Koizumi, Michael Monkenbusch, Dieter Richter, Dietmar Schwahn, and Bela
  Farago.
\newblock Concentration fluctuations in polymer gel investigated by neutron
  scattering: Static inhomogeneity in swollen gel.
\newblock {\em The Journal of chemical physics}, 121(24):12721--12731, 2004.

\bibitem{liang2018nanofishing}
Xiaobin Liang and Ken Nakajima.
\newblock Nanofishing of a single polymer chain: Temperature-induced
  coil--globule transition of poly (n-isopropylacrylamide) chain in water.
\newblock {\em Macromolecular Chemistry and Physics}, 219(3):1700394, 2018.

\bibitem{koizumi2019necklace}
Satoshi Koizumi, Masahiko Annaka, and Dietmar Schwahn.
\newblock Necklace-like microstructure in shallow-quenched aqueous solutions of
  poly (n-isopropylacrylamide), detected by advanced small-angle neutron
  scattering methods.
\newblock {\em Soft matter}, 15(4):671--682, 2019.

\bibitem{futscher2017role}
Moritz~H Futscher, Martine Philipp, Peter M{\"u}ller-Buschbaum, and Alfons
  Schulte.
\newblock The role of backbone hydration of poly (n-isopropyl acrylamide)
  across the volume phase transition compared to its monomer.
\newblock {\em Scientific reports}, 7(1):1--10, 2017.

\bibitem{kang2016collapse}
Yunwon Kang, Heesun Joo, and Jun~Soo Kim.
\newblock Collapse--swelling transitions of a thermoresponsive, single poly
  (n-isopropylacrylamide) chain in water.
\newblock {\em The Journal of Physical Chemistry B}, 120(51):13184--13192,
  2016.

\bibitem{liu2022coil}
Jianyu Liu, Huazhang Guo, Qingjie Gao, Hongbin Li, Zesheng An, and Wenke Zhang.
\newblock Coil--globule transition of a water-soluble polymer.
\newblock {\em Macromolecules}, 55(19):8524--8532, 2022.

\bibitem{huggins2016studying}
David~J Huggins.
\newblock Studying the role of cooperative hydration in stabilizing folded
  protein states.
\newblock {\em Journal of structural biology}, 196(3):394--406, 2016.

\bibitem{miyawaki2016cooperative}
Osato Miyawaki, Michiko Dozen, and Kaede Hirota.
\newblock Cooperative hydration effect causes thermal unfolding of proteins and
  water activity plays a key role in protein stability in solutions.
\newblock {\em Journal of bioscience and bioengineering}, 122(2):203--207,
  2016.

\bibitem{metropolis1953equation}
Nicholas Metropolis, Arianna~W Rosenbluth, Marshall~N Rosenbluth, Augusta~H
  Teller, and Edward Teller.
\newblock Equation of state calculations by fast computing machines.
\newblock {\em The journal of chemical physics}, 21(6):1087--1092, 1953.

\bibitem{harris1988lattice}
Jonathan Harris and Stuart~A Rice.
\newblock A lattice model of a supported monolayer of amphiphile molecules:
  Monte carlo simulations.
\newblock {\em The Journal of chemical physics}, 88(2):1298--1306, 1988.

\bibitem{frenkel1992novel}
D~Frenkel, GCAM Mooij, and B~Smit.
\newblock Novel scheme to study structural and thermal properties of
  continuously deformable molecules.
\newblock {\em Journal of Physics: Condensed Matter}, 4(12):3053, 1992.

\bibitem{siepmann1992configurational}
J{\"o}rn~Ilja Siepmann and Daan Frenkel.
\newblock Configurational bias monte carlo: a new sampling scheme for flexible
  chains.
\newblock {\em Molecular Physics}, 75(1):59--70, 1992.

\bibitem{frenkel2023understanding}
Daan Frenkel and Berend Smit.
\newblock {\em Understanding molecular simulation: from algorithms to
  applications}.
\newblock Elsevier, 2023.

\bibitem{fredrickson2006equilibrium}
Glenn Fredrickson.
\newblock {\em The equilibrium theory of inhomogeneous polymers}.
\newblock Number 134. Oxford University Press, 2006.

\bibitem{dahanayake2021hydrogen}
Rasika Dahanayake and Elena~E Dormidontova.
\newblock Hydrogen bonding sequence directed coil-globule transition in water
  soluble thermoresponsive polymers.
\newblock {\em Physical review letters}, 127(16):167801, 2021.

\bibitem{huang2001dynamic}
Jianhua Huang, Wenhua Jiang, and Shijun Han.
\newblock Dynamic monte carlo simulation on the polymer chain with one end
  grafted on a flat surface.
\newblock {\em Macromolecular theory and simulations}, 10(4):339--342, 2001.

\bibitem{zhang2019modeling}
Yuchong Zhang and Walter~G Chapman.
\newblock Modeling lower critical solution temperature behavior of associating
  dendrimers using density functional theory.
\newblock {\em Langmuir}, 35(33):10808--10817, 2019.

\bibitem{de2018molecular}
Tiago~E de~Oliveira, Carlos~M Marques, and Paulo~A Netz.
\newblock Molecular dynamics study of the lcst transition in aqueous poly
  (nn-propylacrylamide).
\newblock {\em Physical Chemistry Chemical Physics}, 20(15):10100--10107, 2018.

\bibitem{gong2013modeling}
Kai Gong, Bennett~D Marshall, and Walter~G Chapman.
\newblock Modeling lower critical solution temperature behavior of associating
  polymer brushes with classical density functional theory.
\newblock {\em The Journal of chemical physics}, 139(9):094904, 2013.

\bibitem{furyk2006effects}
Steven Furyk, Yanjie Zhang, Denisse Ortiz-Acosta, Paul~S Cremer, and David~E
  Bergbreiter.
\newblock Effects of end group polarity and molecular weight on the lower
  critical solution temperature of poly (n-isopropylacrylamide).
\newblock {\em Journal of Polymer Science Part A: Polymer Chemistry},
  44(4):1492--1501, 2006.

\bibitem{ieong2011missing}
Nga~Sze Ieong, Martin Redhead, Cynthia Bosquillon, Cameron Alexander, Malcolm
  Kelland, and Rachel~K O’Reilly.
\newblock The missing lactam-thermoresponsive and biocompatible poly
  (n-vinylpiperidone) polymers by xanthate-mediated raft polymerization.
\newblock {\em Macromolecules}, 44(4):886--893, 2011.

\bibitem{wu1998globule}
Chi Wu and Xiaohui Wang.
\newblock Globule-to-coil transition of a single homopolymer chain in solution.
\newblock {\em Physical review letters}, 80(18):4092, 1998.

\bibitem{ray2005effect}
Biswajit Ray, Yoshio Okamoto, Masami Kamigaito, Mitsuo Sawamoto, Ken-ichi Seno,
  Shokyoku Kanaoka, and Sadahito Aoshima.
\newblock Effect of tacticity of poly (n-isopropylacrylamide) on the phase
  separation temperature of its aqueous solutions.
\newblock {\em Polymer journal}, 37(3):234--237, 2005.

\bibitem{hirano2006dual}
Tomohiro Hirano, Yuya Okumura, Hiroko Kitajima, Makiko Seno, and Tsuneyuki
  Sato.
\newblock Dual roles of alkyl alcohols as syndiotactic-specificity inducers and
  accelerators in the radical polymerization of n-isopropylacrylamide and some
  properties of syndiotactic poly (n-isopropylacrylamide).
\newblock {\em Journal of Polymer Science Part A: Polymer Chemistry},
  44(15):4450--4460, 2006.

\bibitem{mukherji2017reply}
Debashish Mukherji, Manfred Wagner, Mark~D Watson, Svenja Winzen, Tiago~E
  de~Oliveira, Carlos~M Marques, and Kurt Kremer.
\newblock Reply to the ‘comment on “relating side chain organization of
  pnipam with its conformation in aqueous methanol”’by n. van der vegt and
  f. rodriguez-ropero, soft matter, 2017, 13.
\newblock {\em Soft Matter}, 13(12):2292--2294, 2017.

\bibitem{maffi2012first}
Carlo Maffi, Marco Baiesi, Lapo Casetti, Francesco Piazza, and Paolo
  De~Los~Rios.
\newblock First-order coil-globule transition driven by vibrational entropy.
\newblock {\em Nature Communications}, 3(1):1--8, 2012.

\bibitem{schilling1990longest}
Mark~F Schilling.
\newblock The longest run of heads.
\newblock {\em The College Mathematics Journal}, 21(3):196--207, 1990.

\bibitem{swendsen1986replica}
Robert~H Swendsen and Jian-Sheng Wang.
\newblock Replica monte carlo simulation of spin-glasses.
\newblock {\em Physical review letters}, 57(21):2607, 1986.

\bibitem{sugita1999replica}
Yuji Sugita and Yuko Okamoto.
\newblock Replica-exchange molecular dynamics method for protein folding.
\newblock {\em Chemical physics letters}, 314(1-2):141--151, 1999.

\bibitem{allen2017computer}
Michael~P Allen and Dominic~J Tildesley.
\newblock {\em Computer simulation of liquids}.
\newblock Oxford university press, 2017.

\bibitem{kumar1992weighted}
Shankar Kumar, John~M Rosenberg, Djamal Bouzida, Robert~H Swendsen, and Peter~A
  Kollman.
\newblock The weighted histogram analysis method for free-energy calculations
  on biomolecules. i. the method.
\newblock {\em Journal of computational chemistry}, 13(8):1011--1021, 1992.

\end{thebibliography}

\end{document}